\documentclass{aa}  

\usepackage{graphicx}
\usepackage{txfonts}
\usepackage{natbib}
\bibpunct{(}{)}{;}{a}{}{,}

\def\deg{$^{\rm o}$}
\def\arcmin{$^{\prime}$}
\def\arcsec{$^{\prime\prime}$}

\begin{document}

   \title{The Sardinia Radio Telescope: }

   \subtitle{From a Technological Project to a Radio Observatory }

   \author{I. Prandoni\inst{1}\fnmsep\thanks{\email{prandoni@ira.inaf.it}}
          \and M. Murgia \inst{2}
          \and A. Tarchi \inst{2}
          \and M. Burgay\inst{2} \and P. Castangia\inst{2} \and E. Egron\inst{2} \and F. Govoni\inst{2} \and A. Pellizzoni\inst{2} \and R. Ricci\inst{2} \and S. Righini\inst{1} \and M. Bartolini\inst{1} \and S. Casu\inst{2} \and A. Corongiu\inst{2} \and M.~N. Iacolina\inst{2} \and A. Melis\inst{2} \and F.~T. Nasir\inst{2}  \and A. Orlati\inst{1} \and D. Perrodin\inst{2}  \and S. Poppi\inst{2}  \and A. Trois\inst{2} \and V. Vacca\inst{2}  \and A. Zanichelli\inst{1} \and M. Bachetti\inst{2} \and M. Buttu\inst{2} \and G. Comoretto\inst{3} \and R. Concu\inst{2} \and A. Fara\inst{2} \and F. Gaudiomonte\inst{2}  \and F. Loi\inst{2%,4
          } \and C. Migoni\inst{2}  \and A. Orfei\inst{1}  \and M. Pilia\inst{2}   \and P. Bolli\inst{3}  \and E. Carretti\inst{2}   \and N. D'Amico\inst{2}  \and D. Guidetti\inst{1} \and S. Loru\inst{2} \and F. Massi\inst{3} \and T. Pisanu\inst{2}  \and I. Porceddu\inst{2} % \and A. Possenti\inst{2} 
          \and A. Ridolfi\inst{2} \and G. Serra\inst{2} \and C. Stanghellini\inst{1}   \and C. Tiburzi\inst{2} \and S. Tingay\inst{1} \and G. Valente\inst{2} \\ 
          }

   \institute{INAF-Istituto di Radioastronomia, 
              Via P. Gobetti 101, I-40129 Bologna\\
         \and
             INAF-Osservatorio Astronomico di Cagliari,  Via della Scienza 5, I-09047 Selargius \\
         \and
             INAF-Osservatorio Astronomico di Arcetri,  Largo E. Fermi 5, I-50125 Firenze \\
}

   \date{Received 14 December 2016; accepted 27 March 2017}

  \abstract
   {The Sardinia Radio Telescope (SRT) is the new 64-m dish operated by INAF (Italy).  Its active surface, made up of 1008 separate aluminium panels supported by electromechanical actuators, will allow us to observe at frequencies of up to 116~GHz. At the moment, three receivers, one per focal position, have been installed and tested: a 7-beam K-band receiver; a mono-feed C-band receiver; and a coaxial dual-feed L/P band receiver. The SRT was officially opened in October 2013, upon completion of its technical commissioning phase. In this paper, we provide an overview of the main science drivers for the SRT, describe the main outcomes from the scientific commissioning of the telescope, and discuss a set of observations demonstrating the SRT's scientific capabilities.}
   {The scientific commissioning phase, carried out in the 2012--2015 period, proceeded in stages by following the implementation and/or fine-tuning of advanced sub-systems such as the active surface, the derotator, new releases of the acquisition software, etc. One of the main objectives of scientific commissioning  was the identification of deficiencies in the instrumentation and/or in the telescope sub-systems for further optimization. As a result, the overall telescope performance has been significantly improved. }
   {As part of the scientific commissioning activities, different observing modes were tested and validated, and first astronomical observations were carried out to demonstrate the  science capabilities of the SRT. In addition, we developed astronomer-oriented software tools, to support future observers on-site. In the following, we will refer to the overall scientific commissioning and software development activities  as {\it Astronomical Validation}. } 
   {The astronomical validation activities were prioritized based on technical readiness and scientific impact. The highest priority was to make the SRT available for joint observations as part of European networks. As a result, the SRT started to participate (in 
 shared-risk mode) in EVN (European VLBI Network)  and LEAP (Large European Array for Pulsars) observing sessions in early 2014.  The validation of single-dish operations for the suite of SRT first light receivers and backends continued in the following years, and was concluded with the first call for shared-risk/early-science observations issued at the end of 2015. As discussed in the paper, the SRT capabilities were tested (and optimized when possible) for several different observing modes: imaging; spectroscopy; pulsar timing; and transients.}
   {}


   \maketitle
%
%________________________________________________________________

\section{Introduction}
The Sardinia Radio Telescope\footnote{www.srt.inaf.it}  is a new, general-purpose, fully-steerable 64-m diameter parabolic radio telescope capable of operating with high efficiency in the 0.3-116~GHz frequency range. The instrument is the result of a scientific and technical collaboration among three Structures of the Italian National Institute for Astrophysics (INAF): the Institute of Radio Astronomy (IRA); the Cagliari Astronomical Observatory (OAC); and the Arcetri Astrophysical Observatory (OAA). The main funding agencies are the Italian Ministry of Education and Scientific Research (MIUR), the Sardinia Regional Government (RAS), the Italian Space Agency (ASI), and INAF itself. The SRT is designed to be used for astronomy, geodesy and space science, both as a single dish and as part of European and International networks.
The SRT will operate as an international facility, with regular world-wide 
distributed calls for proposal\footnote{Details about the time allocation procedures and policy will be released at a suitable 
time before the beginning of each observing term.}, and no a priori limitation based on the 
affiliation of the proposers.  A large fraction of the observing time (of the order of 80\%) will be devoted to radio astronomy applications, while 20\% of the time will be allocated to 
activities of interest to the Italian Space Agency (ASI). 

The SRT is located in the plain of Pranu Sanguni (Lat. 39$^{\circ}$29$^{\prime}$34$^{\prime\prime}$N - Long. 9$^{\circ}$14$^{\prime}$4$^{\prime\prime}$ E) at an altitude of 600 m a.s.l., 35~km north of Cagliari (IT), in the municipality of San Basilio. The manufacturing and the on-site assembly of its mechanical parts  were commissioned in 2003 and completed in mid-2012. The antenna officially opened on September 30th 2013, upon completion of the technical commissioning phase (see \citealt{bolli15}). Scientific commissioning of the SRT was carried out in the 2012--2015 period.

The SRT started its scientific operations in 2013, when a Target of Opportunity (ToO)  programme was offered to the community on a best-effort basis. After the first successful VLBI (Very Long Baseline Interferometry) data correlation, which was obtained between the SRT and Medicina stations in January 2014, the SRT has regularly participated in EVN (European VLBI Network) test observations (\citealt{migoni14}), and since 2015, the SRT has been offered as an additional EVN station in shared-risk mode.   
In addition, we have successfully implemented the LEAP (Large European Array for Pulsars) project at the SRT, having installed all of the hardware and software  necessary for the project (\citealt{perrodin14}). Since early 2014, the SRT has participated, together with four other European radio telescopes,  in monthly LEAP runs, for which data acquisition is now fully automated.  At the end of 2015, the first call for single-dish (early science/shared risk) projects with the SRT was issued, and the early science observations started on February 1st, 2016. 

The aim of this paper is to provide a general overview of the science applications planned for the SRT, and to report on the scientific commissioning of the SRT's first-light receivers and backends. The SRT scientific commissioning was carried out in steps, from basic "on-sky" tests aimed at verifying the general performance and/or the limits of the telescope and the acquisition systems, to more complex acquisitions aimed at assessing the actual SRT capabilities  for typical scientific observations. Examples of the former are: the verification of the backend linearity range; the verification of the On-The-Fly (OTF) scan pointing accuracy and maximum exploitable speed; the measurement of the confusion noise; and an accurate characterization of the beam side-lobes. Examples of the latter are continuum and/or spectroscopic acquisitions, including mapping of extended sources, pulsar timing, etc. As part of the scientific commissioning activities, several ad hoc software tools were also developed and implemented at the SRT site in order to provide support for observers.  In the following, we will refer to the overall scientific commissioning and software development activities  as {\it Astronomical Validation (AV)}. This paper summarizes the  various aspects of the AV activities and provides a reference for the SRT scientific performance with the suite of first-light receivers and backends, with particular focus on single-dish operations. We refer to \cite{prandoni14} for a more detailed discussion of  the use of the SRT in the context of VLBI networks.  

This paper is organized as follows. In Sect.~\ref{sec:SRT}, we give  a brief description of the SRT and its current instrumentation, while in 
Sect.~\ref{sec:obs-tools}, we illustrate the available suite of on-site software tools, developed as part of the AV to support SRT observations, data reduction and analysis. An overview of the scientific areas  where the SRT can play an important role is given in Sect.~\ref{sec:science}.  We then illustrate the main results obtained as part of the SRT's scientific commissioning. The AV activities ranged from on-sky tests aimed at testing basic SRT performance (Sects.~\ref{sec:telperf} and~\ref{sec:tpperf}), to more advanced observations aimed at  highlighting the SRT scientific capabilities for  a range of applications: imaging (Sect.~\ref{sec:imaging}); spectroscopy (Sect.~\ref{sec:xarcos}); and pulsar observations (Sect.~\ref{sec:pulsar}). We conclude by reporting on the first results obtained  with the SRT as part of the LEAP network (Sect.~\ref{sec:LEAP}) and for observations of transients as part of the ToO programme (Sect.~\ref{sec:ToO}). A brief summary is provided in Sect.~\ref{sec:summary}. 

Throughout this paper, we generally adopt the \cite{perley13} flux calibration scale.  When specified, the \cite{baars77} scale is used instead. We notice that the measured systematic differences between the two flux scales ($\la 5\%$ for the observed calibrators at the relevant frequencies, see Table 13 of \citealt{perley13}) do not have any significant impact on the reported results of the scientific commissioning.

%__________________________________________________________________
%__________________________________________________ One column table
   \begin{table}
     \centering 
     \caption[]{SRT technical specifications. }\label{tab:SRT}
      \begin{tabular}{@{} l  | l @{}}
            \hline\hline
Parameter & Value or Range \\
          &      \\
          \hline
Elevation range & $5^{\circ} - 90^{\circ}$ \\
Azimuth range & $\pm 270^{\circ}$ \\
Azimuth slewing speed & 51$^{\circ}$/min \\
Elevation slewing speed & 30$^{\circ}$/min \\
Surface accuracy($^{\dag}$) & 305 $\mu$m \\
Pointing accuracy$(^{\ddag})$ & $2/13 $ arcsec  \\
%          &      \\
            \hline
      \end{tabular}   
\tablefoot{
\tablefoottext{$\dag$}{Spec at 45$^{\circ}$ Elevation.}
\tablefoottext{$\ddag$}{Specs at 22 GHz for normal conditions with/without metrology implemented.}
}
   \end{table}

\begin{table*}[]
\centering
\caption{Reference values for relevant parameters of the SRT first-light receivers. 
}
\label{tab:receivers}
\begin{tabular}{lccccccl}
\hline\hline
%         &         &     &    &     &    &      &       \\
Receiver & RF Band & BW$_{\rm max}$ & T$_{\rm Rx}$  & T$_{\rm sys}^{\dag}$ 
& G & HPBW$^{\ddag}$ & Focal  position \& focal ratio ($f/D$) \\
  &(GHz) &(MHz)&  (K)&  (K)& (K/Jy) & (arcmin) &  \\
\hline
% &          &         &     &    &     &    &      &       \\
L/P-band   & 0.305-0.41 (P) & 105  & 20 &  52  & 0.53
    & 55 & Primary: $f/D=0.33$\\
                  &1.3-1.8 (L) & 500 &  11 &  20  & 0.52
                      &  12.5 & \\
\hline
%          &         &     &    &     &    &      &       \\
C-band                 &5.7-7.7  &2000 & 7.7  &  29  & 0.60  & 2.8  & BWG:  $f/D=1.37$\\ 
\hline
%         &         &     &    &     &    &      &       \\
K band     &18-26.5  &2000 & 22 &  75-80  
     & 0.65  
          & 0.82 & Gregorian: $f/D=2.35$\\
%         &         &     &    &     &    &      &        \\
\hline
\end{tabular}
\tablefoot{
\tablefoottext{$\dag$}{System temperature measured at 45$^{\circ}$ elevation. }
\tablefoottext{$\ddag$}{Indicative value at the central frequency of the RF band.}
}
\end{table*}

\section{The SRT in a nutshell}\label{sec:SRT}
A full description of  the SRT telescope is beyond the scope of the present paper. Here we only highlight its main characteristics, and we refer the reader to the SRT technical commissioning paper (\citealt{bolli15}) for more details.  

The SRT has a shaped Gregorian optical configuration with a 7.9-m diameter secondary mirror and
supplementary Beam-Wave-Guide (BWG) mirrors. We currently have four focal positions (primary, Gregorian, and two
BWGs\footnote{In the future, two more BWG focal positions will be implemented for space science applications.}) that allow us to allocate up to 20 remotely-controllable receivers (\citealt{buttu12}). In its first light configuration, the SRT is equipped with three receivers, one per focal position: a 7-beam K-band (18-26 GHz) receiver (Gregorian focus); a mono-feed C-band receiver, centered at the 6.7--GHz methanol line (BWG focal position); and a coaxial dual-feed L/P band receiver, with central frequencies  of 350 MHz and 1.55 GHz, respectively (primary focus). Additional receivers are currently under development: a mono-feed  covering the lower frequency part of the C-band ($\sim 5$ GHz), and two multi-feed receivers (respectively of 7 and 19 beams) operating at S- and Q-bands (i.e. at 3 and 43 GHz). 

One of the most advanced technical features of the SRT is its active surface.  The primary mirror is composed of 1008 panels, which are supported by electromechanical actuators that are digitally controlled to compensate for deformations. 
Table~\ref{tab:SRT} reports the main technical specifications of the telescope. We note that the surface accuracy reported in Table~\ref{tab:SRT} refers to the photogrammetry panel alignment, which is the current implementation of the active surface at the SRT, and which corrects for gravitational deformations only. 
Photogrammetry measurement campaigns allowed the production of a look-up table (LUT), which is used to correct for gravity deformations. The  LUT  provides the translation and rotation corrections to be applied as a function of elevation. This accuracy is appropriate for obtaining a high-efficiency performance for operating frequencies up to $\la 50$ GHz, and is therefore fully suitable for the receiver suite that is currently available. Advanced metrology techniques will be needed to correct for thermal and wind pressure deformations, and to obtain a high efficiency performance up to the maximum frequency for which the SRT is designed to operate ($116$ GHz). The active surface can also be used to re-shape the primary mirror from a {\it shaped} configuration to a parabolic profile, which is recommended for increasing the field of view and the efficiency when using the receivers positioned at the primary focus (\citealt{bolli14}). Table~\ref{tab:receivers} summarizes the main parameters of the three first-light receivers, namely: the radio frequency band covered by the receiver (RF-Band); the maximum instantaneous bandwidth (BW$_{\rm max}$); the receiver temperature (T$_{\rm Rx}$); the system temperature (T$_{\rm sys}$); the gain (G); the beam width at half power (HPBW) of the main lobe of the telescope beam\footnote{HPBW (in arcmin) roughly scales as $k/\nu$, where k is a RF-band dependent constant and  $\nu$ is the frequency (in GHz): $k=19.590$ for P-band;  $k=19.373$ for L-band; 18.937 for C-band; $k=18.264$ for K-band.}; and the focal position (Primary, Gregorian, BWG) at which the receiver is mounted, together with its focal ratio ($f/D$, where $f$ is the focal length and $D$ is the telescope diameter).  We note that on-sky measurements were performed in shaped Gregorian optical configuration, except for L/P-band where the primary mirror was re-configured to a parabolic profile.  The K-band T$_{\rm sys}$ range reported in Tab.~\ref{tab:receivers} reflects a number of measurements in different 2~GHz sub-bands, at various atmospheric opacity values in the range $\tau \sim 0.04-0.06$.  In addition, the listed gain values should be considered as indicative: for K-band, we report an average value in the elevation range of $50^{\circ} - 80^{\circ}$; for P-band, only an expected value is given, as measurements were severely affected by RFI. For constantly updated values, we refer to the SRT website.

The currently-available suite of backends at the SRT is listed below (for more details, we refer to \citealt{melis14c}): 

\begin{description}
 \item  {\it Total Power (TP)}: 14 voltage-to-frequency converters digitize the detected signals. Different Intermediate Frequency (IF)  inputs can be selected from three focal points. The system can be adjusted to select different instantaneous bandwidths (up to a maximum of 2~GHz), and to modify the attenuation level. \\
 
 \item {\it XARCOS}:  narrowband spectro-polarimeter with 16 input channels. The nominal working band is 125-250~MHz. The band that is actually exploitable is 140-220~MHz.  The number of channels provided  for each double-polarization feed is $2048\times 4$ (full Stokes). The instantaneous bandwidth ranges between 0.488 and 62.5~MHz. This means that  the achievable maximum spectral resolution is $\sim 238$~Hz. Up to four tunable sub-bands can be simultaneously employed in conjunction with the mono-feed operating at C-band, as well as with the central feed of the K-band receiver. In the latter case, only two sub-bands can be simultaneously set when performing observations in nodding mode (see \citealt{melis15} for more details). \\

\item {\it Digital Base Band Converter (DBBC)}: digital platform based on a flexible architecture, composed of four Analog-to-Digital Converter (ADC) boards, 1~GHz bandwidth each, and four Xilinx Field-Programmable Gate Array (FPGA) boards for data processing. The DBBC platform is designed mainly for VLBI experiments.  However, a different firmware allowing wide-band spectrometry has been developed for different purposes, such as the monitoring of Radio Frequency Interference (RFI, see Sect.~\ref{sec:obs-tools}). \\

\item {\it Digital Filter Bank Mark 3 (DFB3)}: FX correlator developed by the Australia Telescope National Facility, allowing full-Stokes observations. It has four inputs with a 1024~MHz maximum bandwidth each, and 8-bit sampling for a high dynamic range. The DFB3 is suitable for precise pulsar timing and searching, spectral line, and continuum observations with a high time resolution. It allows up to 8192 spectral channels, to counter the effects of interstellar dispersion when operated in pulsar mode, and for power spectrum measurements in spectrometer mode.\\
    
\item   {\it ROACH}: digital board developed by CASPER\footnote{Collaboration for Astronomy Signal Processing and Electronics Research.}. The centerpiece of the board is a Xilink Virtex 5 FPGA. It is currently configured with a personality which provides 32 complex channels of 16 MHz each (for a total bandwidth of 512 MHz). This backend is adopted for pulsar observations in the context of LEAP.\\

\item  {\it SARDARA}: The SArdinia Roach2-based Digital Architecture for Radio Astronomy (SARDARA) is a new, wide-band multi-feed digital backend that is currently under development; it will be exploitable for both continuum studies and as a full-Stokes spectrometer. The FPGA-based ROACH2 boards are used as the main processing cores, together with an infrastructure including GPU-based nodes, a 10--Gb Ethernet SFP+  (Small Form-factor Pluggable) switch and a powerful storage computer (\citealt{melis16}). A preliminary version of SARDARA, only suitable for single-feed double-polarization receivers, is already available at the telescope (see the SRT website for updates on supported configurations).

\end{description}

Both TP and XARCOS are designed to exploit the multi-feed receiver operating at K-band (7 feeds $\times 2$ polarizations, i.e. 14 output channels), but can also be used for observations at C-band. The current Local Oscillator setup does not allow us to use XARCOS at P- and L-bands. The use of the Total Power analog backend, although in principle allowed,  is not recommended at P- and L-bands due to severe RFI pollution, which limits its performance (the RFI environment at the SRT site is discussed extensively in \citealt{bolli15}). 
In this paper, we will therefore focus on C- and K-bands observations with the Total Power and  XARCOS backends.  

For lower frequency observations, we will rely on the DFB3 and ROACH backends, which are mainly used for pulsar science applications. The use of the DBBC in the context of VLBI networks is illustrated in \cite{prandoni14} and will not be discussed here. For the scientific demonstration of SARDARA (both for continuum and polarization observations), we refer to \cite{murgia16} and \citet{melis16}.

The telescope is managed by means of a dedicated control system called {\it Nuraghe} (\citealt{orlati12}), which was developed within the DISCOS project (\citealt{orlati15}),  and which provides all of the INAF radio telescopes with an almost identical  common control system. {\it Nuraghe} is a distributed system that was developed using the ALMA Common Software (ACS) framework. It handles all of the operations of the telescope, taking care of the major and minor servo motions, the frontend setup, and the data acquisition performed with the integrated backends - at present, the TP and XARCOS - and producing standard FITS output files. Additional backends can be used in semi-integrated or external modes. {\it Nuraghe} also reads the LUT and commands the 6 electro-mechanical actuators to correct for both the optical misalignments and the primary mirror deformations due to gravitational effects. The user interface allows the real-time monitoring of all the telescope devices. Automatic procedures let the user easily carry out essential operations like focusing, pointing calibration, or skydip scanning. {\it Nuraghe} supports observing modes such as sidereal tracking, On-Off position switching\footnote{On-Off position switching is an observing mode typically used for spectroscopic observations. It corresponds to alternate on-source/off-source pointing sequences. The off-source spectrum is used to remove atmospheric and system noise, and to bandpass calibrate the on-source spectrum.}, On-The-Fly (OTF) cross-scans\footnote{On-the-Fly cross-scans (or simply cross-scans) are fast acquisitions carried out along two orthogonal directions, usually coinciding with the axes of a celestial coordinate frame (Az-El, RA-Dec, GLon-GLat). They are mainly employed to observe point-like sources, in order to ultimately measure their flux density and position. Cross-scans allow an accurate estimate of the source peak position, in turn permitting us to correct for pointing errors (local residuals with respect to the pointing model).}, raster scans\footnote{Raster scans are sequences of pointed acquisitions organized in grids, mainly used to sample discrete portions of extended targets according to a pre-defined geometry.}, and OTF mapping\footnote{OTF mapping is an observing mode exploiting fast OTF scans across a given area. Scans are properly weaved and interspaced, in order to achieve the requested sampling in the final map.} in the equatorial, galactic and horizontal coordinate frames. For details and instructions, we refer to the official documentation, which is available at the DISCOS project website\footnote{{\it http://discos.readthedocs.org}}. 

\section{The SRT as an astronomical observatory}
\label{sec:obs-tools}

As part of the observatory activities, a list of sources has been monitored at the SRT since the beginning of the AV activities. These observations are aimed at producing a list of validated flux and pointing calibrators for the SRT, as well as  verifying the pointing model accuracy on a regular basis. 
A list of validated calibrators for C- and K-bands observations with the  SRT is available in \cite{ricci16}.  Updates will be advertised on the SRT website. 

The AV included a preparatory phase that was carried out during technical commissioning of the telescope. Several external software (SW) tools were developed in order to assist  the preparation, execution, and monitoring of the observations, as well as the data inspection and reduction. Some of these tools were made publicly available online, while some other tools are available to observers on-site or meant to support the observatory personnel. Below we give a brief description of the main developed tools.  For details and updates on the released versions of these tools, we refer to the SRT website. We note that these tools were all extensively tested and used for the AV activities (observations, data reduction and analysis). Whenever relevant, their performance and capabilities are further discussed in more detail in the following sections.

\subsection{Preparing the observations}\label{sec:tools1}

\begin{description}
\item {\it ETC:} The Exposure Time Calculator provides an estimate of the exposure time needed to reach a given sensitivity (or vice versa) under a set of assumptions on the telescope setup and the observing conditions. In its current version, it includes both the SRT and Medicina telescopes (Noto will be added in the near future). Details and instructions are provided in a dedicated user manual (\citealt{zanichelli15}).\\
\item {\it CASTIA\footnote{CASTIA means "look" in Sardinian.}:}  this software package provides radio source visibility information at user-selected dates for any of the three INAF radio telescope sites (SRT, Medicina and Noto), as well as for a collection of more than 30 international radio facilities. The tool produces plots showing the visibility (and the elevation) of radio-sources versus time, highlighting the rise, transit, and set times. Warnings are provided when the Azimuth rate is beyond the recommended limit and when superposition with the Sun and/or the Moon occurs. For a detailed description of the tool and of its usage, we refer to \cite{vacca13}.\\
\item {\it Meteo Forecasting:} this tool uses a numerical weather prediction model on timescales of 36 hours, to allow for dynamic scheduling (for more details, see \citealt{nasir13,buffa16}).\\
\item {\it ScheduleCreator:} this tool produces properly-formatted schedules for {\it Nuraghe}, the telescope control software, for all available observing modes (sidereal tracking, On-Off, OTF cross-scans, raster scans, and mapping in the equatorial, galactic and horizontal coordinate frames).  The target list and information on both the system setup and the execution of the observations are given as input parameters. Details and instructions are provided in a dedicated user manual (\citealt{bartolini13,bartolini16}). 
\end{description}

\subsection{Executing  and monitoring the observations}\label{sec:tools2}
As discussed in Sect.~\ref{sec:SRT}, the antenna control software {\it Nuraghe} allows the user to execute continuum and spectroscopy observations with the fully integrated backends (TP and XARCOS), as well as monitor all of the different telescope sub-systems (antenna mount, active surface, front-ends, etc.). As part of the AV, an additional dedicated control software has been developed for managing pulsar observations. This software is called SEADAS (SRT ExpAnded Data Acquisition System). In addition, a number of other software tools have been developed to assist the observers in data quality monitoring during the observations. Brief descriptions of SEADAS and the other monitoring tools are provided below. 

\begin{description}
\item {\it SEADAS} communicates directly with {\it Nuraghe} for the purpose of antenna configuration and pointing, while also interacting with specifically designed software tools running on the backend's server for the purpose of setting up and coordinating the data acquisition.  Particular attention has been given to the schedule tool, which allows the user to easily read and edit the schedules, and to the organization of the graphic interface, which permits the straightforward monitoring of the whole observing session. SEADAS has been designed to allow data acquisition with multiple backends in parallel. The DFB3 is the first backend, and to date the only one, whose control is fully integrated in SEADAS.  The integration of the ROACH backend is currently under development and will be effective in the near future.  On a longer timescale, all present (and future) backends for pulsar observations will be integrated in SEADAS. The documentation for preparing and running a pulsar observing session with SEADAS is maintained on the SRT website, while for a full description of SEADAS, we refer to  \cite{corongiu14}.  \\
\item {\it FITS Quick Look:} this IDL program can handle the quasi real-time display of both mono-feed and multi-feed data.  The present release displays the content of FITS files acquired using the Total Power backend only (spectrometry FITS will be added in the near future). Data streams can be shown as raw counts or antenna temperature vs. time or celestial coordinates. \\
\item {\it RFI Monitoring:}  a piggy-back RFI monitoring system uses the DBBC, during observations with other backends. The infrastructure consists essentially of a wide-band Fast Fourier Transform (FFT)  spectrometer operating on a copy of the radioastronomical signal, and a Linux-based PC containing an RFI detection pipeline (\citealt{melis14a,melis14b}).
\end{description}

\subsection{Data Reduction Tools}\label{sec:tools3}

\begin{description}
\item {\it RFI Flagger:}  a tool developed to create flag masks of spectroscopic data (e.g. from SARDARA), based on frequency- and time-domain algorithms (Ricci et al., in prep.). \\
\item {\it Output file converters:} tools that allow the conversion of the SRT Fitzilla output files to other FITS formats (e.g. SDFITS and the input FITS format for the GILDAS data reduction package; Trois et al. in prep.). \\
\item {\it Cross Scans}: data reduction software for the integration and calibration of continuum cross-scans acquired on point-like sources. \\
\item {\it SDI:} SRT Single-Dish Imager, a data reduction software package customized for the SRT and the Medicina antennas, which can be used to produce images from OTF scans obtained with either mono- or multi-feed receivers. Its main features are: 1) real-time imaging, through automatic baseline subtraction and RFI flagging; 2) state-of-the-art calibration procedures and data handling tools for further flagging of the data; and 3) standard DS9 FITS output for image inspection and analysis (Pellizzoni et al., in prep.). \\
\item {\it SCUBE:} SCUBE (Single-dish Spectro-polarimetry Software) is a proprietary data reduction software optimized for single-dish spectro-polarimetry data (\citealt{murgia16}). The tool can manage data obtained with the total power backend as well as data obtained with digital backends (e.g. XARCOS, SARDARA), both in total intensity and polarization modes. SCUBE is composed of a series of routines (written in the C++ language) that perform all calculation steps needed to pass from a raw dataset to a final calibrated image. The output FITS files and tables can be analyzed and displayed with the standard FITSViewer programs and graphics packages. \\
\end{description}

\section{Key Science with the SRT}\label{sec:science}

The SRT is a general-purpose facility which was designed for astronomy, geodesy, and space science applications. Thanks to its large aperture and versatility (multi-frequency agility and wide frequency coverage), we expect to make use of the SRT for a wide range of scientific topics for many years to come. Here we highlight some areas where we believe the SRT can play a major role in the next future, both as part of observing networks and as a single dish. We note that operations in the framework of international VLBI and Pulsar Timing networks are top priorities for the SRT.

\subsection{VLBI Science}
The SRT is  one of the most sensitive European VLBI Network (EVN) stations, together with Effelsberg and Jodrell Bank. Its large aperture is also of extreme importance for Space VLBI observations with RadioAstron. Thanks to its active surface, the SRT can also constitute a sensitive element of the mm-VLBI network operating in the 7- and 3-mm bands. At these frequencies, substantial improvements in collecting area and coverage of the sky Fourier transform plane are of vital importance for  either increasing the number of targets that are accessible to the array, or improving the quality of the images. Once the fiber
optic connection to the site is completed, the SRT will also participate in real-time VLBI observations (e-VLBI). The availability of three antennas in Italy allows the constitution of a small independent Italian VLBI network, exploiting a software correlator that is already operating (DiFX, \citealt{deller07}). As soon as it is equipped with the appropriate receivers, the SRT will also be included in geodetic VLBI networks.
The geographical position and large aperture of the SRT are of particular interest for high resolution observations of sources at intermediate declinations (which cannot be studied well by any of the existing VLBI networks). Such observations will become possible in the future using the Italian antennas jointly with the African VLBI Network (AVN), which is currently under development.

\subsection{Gravitational wave detection experiments}
The SRT is one of the five telescopes of the European Pulsar Timing Array (EPTA, \citealt{kramer13}), which also includes Effelsberg, Jodrell Bank, Nancay and the Westerbork Radio Telescope (WSRT). Together with the North American Nanohertz Observatory for Gravitational Waves (NANOGrav, \citealt{hobbs13}), and the Australian Parkes Pulsar Timing Array (PPTA, \citealt{mclaughlin13}) (the three collaborations together form the International Pulsar Timing Array or IPTA, \citealt{manchester13}), the EPTA shares the goal of detecting gravitational waves (GW) at nanohertz frequencies, such as those emitted by supermassive black hole binaries or cosmic strings, using the high precision timing of millisecond pulsars (MSPs). The frequency range of the gravitational waves detectable with pulsar timing arrays is complementary to the frequencies detectable by the current ground-based interferometers, such as LIGO and VIRGO, and by the future space-based interferometer eLISA. Since the SRT is the southernmost telescope of the EPTA collaboration, it will allow a better coverage of pulsars with declinations below $-20$ deg, hence providing a better overlap with the PPTA. Thanks to its dual-band L/P receiver, the SRT will be of great importance in measuring accurate dispersion measure variations, which are crucial to obtaining high precision pulsar timing data, and searching for signatures of space-time perturbations in pulsar timing residuals. The SRT is also part of LEAP (\citealt{bassa16}), an EPTA project which uses the EPTA telescopes in tied-array mode, performing simultaneous observations of MSPs, and obtaining in this way a sensitivity equivalent to that of a fully steerable 200-m dish.

\subsection{Pulsar studies}
Besides its role in high precision pulsar timing in the context of the
aforementioned multi-telescope projects
(both for the detection of GWs and for tests of theories of gravity), the SRT also has great potential for other types of pulsar studies.

The coaxial L/P band receiver is a unique instrument for 
studying {\it eclipsing pulsars}, which are binary neutron stars whose radio signals 
undergo distortions along their orbit as a consequence 
of the interaction with the plasma released by the companion stars.

The intensity and duration of eclipses at different frequencies,
as well as the delays in the times of arrival of the pulses close to the eclipse 
in different bands, encode a wealth of information about binary pulsars, 
such as the geometry of the binary system, the distribution of the plasma 
poured into the binary, and the density and temperature of the plasma. From 
the measurement of these parameters, one can, for instance, constrain the 
timescale for the complete ablation of the companion, and thus support or 
contrast the hypothesis that these systems will give birth to isolated 
MSPs. In addition, such observations, when coupled with radio timing (which provides a description of the 
spin and orbital evolution of the pulsar), allow us to determine 
the amount of angular momentum loss per unit mass, a 
fundamental quantity for predicting the long term orbital evolution of 
these binaries.

A simultaneous dual-frequency study is also fundamental for clarifying the 
physical mechanisms underlying the eclipses, for which several competing 
models have been proposed (e.g. \citealt{phinney88,rasio89,rasio91,stappers01,gedalin93,thompson94,khechinashvili00}).

The SRT can also be used for surveying the sky in the search for new pulsars: its
high frequency receivers (such as the K-band multi-feed receiver) will be used for targeted searches of the 
Galactic Centre, where the discovery of even a small number of objects that are
gravitationally bound to the central Black Hole would be of paramount 
importance. Larger scale surveys, aimed at increasing the pulsar population 
for statistical purposes and at finding new peculiar pulsars that will be useful for 
precision pulsar timing experiments, are better suited for SRT's lower 
frequency receivers. Besides the single-beam L/P band receiver, the multi-feed
S-band receiver, which is currently under development (see Sect.~\ref{sec:SRT}), will be used to search for both new pulsars and 
fast radio transients (see below).

\subsection{The transient sky}
The SRT can play an important role, either as a single dish or as a sensitive element of VLBI networks, in studying the transient sky. High-frequency receivers will allow us to conduct (high spatial resolution) follow-up observations and monitoring experiments of Active Galactic Nuclei (AGN), Gamma Ray Bursts (GRB), and other transient events, in connection with high-energy experiments (FERMI, MAGIC, CTA). The Italian astronomical community is very active in this research field. A Target of Opportunity programme  has been established at the SRT since 2013. Agreements have  been signed in the framework of international collaborations for the purpose of following up transient events, including Fast Radio Bursts (FRB) and gravitational wave detections. 

\subsubsection{Fast Radio Transients}

The transient radio sky at very short timescales (milliseconds) has begun 
only recently to be systematically explored. Searches for single 
de-dispersed radio pulses of short duration led in recent years to 
the discovery of two peculiar types of sources: Rotating RAdio 
Transients (RRATs; \citealt{mclaughlin06}), which are
rotational-powered neutron stars that, for reasons yet to be clarified, 
emit only sporadic single radio pulses and, more recently, FRBs, which constitute a very intriguing class of still-mysterious short-duration 
radio signals whose dispersion measure allows us to likely place them 
at cosmological distances (\citealt{thornton13,chatterjee17}).

Finding more FRBs is of paramount importance for clarifying their nature 
and for exploiting them as cosmological probes. Every observation made at 
the SRT with the DFB, SARDARA, or the LEAP ROACH 
backends can be simultaneously searched for FRBs in real time, allowing us to 
trigger observations at other wavelengths and to pinpoint the source that is
responsible for the emission of these very peculiar systems. Following up on newly-discovered FRBs is also of great interest, since it can allow us to determine whether these sources, or a sub-class of them, are repeating (like e.g. FRB121102; \citealt{spitler16}), or if they are single signals from disruptive events.

\subsubsection{Electromagnetic Counterparts of GWs}

The detection of GWs by the LIGO-Virgo 
collaboration in 2015 (\citealt{abbott16}) 
has been a milestone discovery, and has made 
possible a direct test of one of the key predictions of general relativity. 
Future prospects are even more exciting: GWs do not experience some of the 
limitations affecting electromagnetic waves, such as the 
absorption and distortion of the signal along the path from the source to the 
observer. Hence, GWs are unique messengers for the physical processes 
occurring in the often unaccessible inner regions of the emitters, provided 
their electromagnetic counterpart(s) are identified. In this context, the SRT can 
play a relevant role as part of a large multi-wavelength program 
(established through a formal agreement), which is aimed at securely identifying and 
following up electromagnetic counterparts of GW events.

\subsubsection{X-ray Binaries}
Radio observations are particularly important for studying
accretion. The radio emission in accreting systems is dominated by the synchrotron
self-Compton emission from a jet that produces a continuum spectrum extending from
radio to X-ray wavelengths. A correlation between X-ray and radio luminosities is
observed in accreting black holes, with masses spanning more than 6 orders of
magnitude (\citealt{merloni03}). X-ray monitoring campaigns of stellar-mass black
hole binaries show a variety of spectral states corresponding to different
accretion regimes, on timescales of a few months. Simultaneous X-ray and radio
campaigns have highlighted corresponding changes in jet power and configuration,
ranging from steady/weak to compact/powerful jets (\citealt{fender14}). This
multi-epoch, multi-wavelength approach allows us to investigate the conditions that
lead to the formation of jets, whose origin is still unclear. 

Only very large-aperture telescopes, such as the SRT, have the sensitivity required to detect faint (mJy) radio emission from jets on the short variability time-scales of this kind of objects (likely of the order of seconds/minutes, or even less).  In addition, the frequency agility implemented at the SRT allows the fast switching from one observing frequency to another, which is crucial to constraining the intrinsic radio spectrum on timescales relevant to variability, including time-dependent departures from the typical flat slope of radio jets.

\subsection{High-frequency Galactic and extra-galactic surveys}
Thanks to its active surface, the SRT can operate with high efficiency at high radio frequencies. The combination of relatively smaller aperture (64-m) and availability of  multi-feed receivers makes the SRT a fast mapping machine. The SRT  can reach $10\times$ larger mapping speeds than its main competitors (Effelsberg and Green Bank, 100-m),  and can therefore play a major role in conducting wide-area surveys of the sky in a frequency range ($20-90$ GHz) that is poorly explored, yet is very interesting. Spectroscopic surveys will in particular benefit from the SRT's shaped Gregorian optical configuration, which mitigates the well-known problem of standing waves\footnote{The phenomenon of {\it standing  waves}  is  caused  by multiple internal reflections between the primary and the secondary mirrors. Standing waves produce periodic ripples in the observed spectra and are particularly detrimental to wide-band spectroscopic observations. A {\it shaped} Gregorian optical configuration produces a null field in the region blocked by the sub-reflector, significantly mitigating this phenomenon. }.

The first-light, K-band 7-beam receiver coupled with the XARCOS spectro-polarimeter can be exploited to map (molecular) spectral lines both in the Milky Way and
in external galaxies. In particular, extensive mapping of the ammonia molecule, in close synergy with existing IR/sub-mm continuum surveys of the Galactic Plane, will provide relevant clues on the physical conditions of the gas in Galactic star-forming regions. Ammonia is considered to be a very good tracer of dense pre-stellar cores, and an excellent thermometer. XARCOS can simultaneously observe its two main inversion transitions, $(1,1)$ and $(2,2)$ at $\sim 23.7$ GHz,  and derive very reliable gas kinetic temperatures.

In addition to thermal lines, the K-band multi-feed receiver can be exploited for extensive searches of H$_2$O maser lines in nearby, spatially-extended galaxies, such as those belonging to the Local Group. Water masers represent a unique tool for deriving, through single-dish and VLBI follow-up monitoring, 3D motions and
distance measurements, ultimately  leading to dynamical models and total mass estimates of (luminous $+$ dark) matter for such galaxies (\citealt{brunthaler05,brunthaler07}). 

Wideband multi-feed receivers operating at higher frequencies ($40 - 90$ GHz) - the one operating at 40 GHz (19 feeds) is currently under development -  will allow us to get access to unique molecular line transitions in our own Galaxy. For example, the transitions associated with deuterated molecules (e.g. DCO$+(1,0)$, N2D$+(1,0)$) are crucial to constraining the kinematic and chemical properties of pre-stellar cores, as well as to uncovering the cool molecular content of the Universe in a crucial cosmic interval (redshift $0.3-2$), through the mapping of redshifted CO $low-J$ transitions.

Wide-band total intensity  and polarization  surveys of the Northern Hemisphere at mid and high frequencies will in turn allow us to obtain important information on the ongoing physical processes in the  radio source populations dominating the sky at such frequencies. 
Even more importantly, these sources also play a vital role in the interpretation of temperature and polarization maps of the cosmic microwave background (CMB). The knowledge of their positions and of their (continuum and polarized) flux densities is crucial to removing their contribution and to estimating the residual error due to faint and unresolved components in CMB maps. 

A first pilot radio continuum survey at 20 GHz was successfully conducted using the Medicina telescope in the early stages of the scientific commissioning activities, and was mainly aimed at validating the K-band multi-feed receiver (\citealt{righini12,ricci13}). This survey covered the Northern sky at $\delta > +72$\deg, down to a flux density limit of $\sim 100$ mJy, and with an angular resolution ($\sim 1.5$ arcmin) similar to the Australia Telescope 20 GHz survey (AT20G; \citealt{murphy10,massardi11}) covering the southern sky. With the multi-feed S-band receiver which will soon be in operation at the SRT, a survey similar to the S-PASS (S-band Polarization All-Sky Survey; \citealt{carretti13}), conducted in the Southern Hemisphere with Parkes, will become feasible with the SRT. 

Finally, wide-band mapping of extended (low-surface brightness) Galactic and extra-galactic sources (e.g. supernova remnants, radio galaxies, nearby spirals) will permit resolved studies (both in radio-continuum and polarization) aimed at a better understanding of the physics of accretion and star formation processes in such sources. Below we provide some examples based on research fields where the Italian community is very active.

\subsubsection{Supernova Remnants}

Observations of Supernova Remnants (SNRs) are a powerful tool for investigating the later stages of stellar evolution, the properties of the ambient interstellar medium, the physics of particle acceleration and shocks, and the origin of Galactic cosmic rays. The multi-wavelength spectrum of SNRs typically feature synchrotron emission, mostly from radio-emitting electrons, and high-energy emission arising from bremsstrahlung and Inverse Compton (IC) processes  produced by the radio-electrons interacting with ambient photons, or hadronic emission provided by $\pi^0$ mesons decay. The long quest for the firm disentanglement among these two scenarios (leptonic versus hadronic models) represents one of the most important challenges for the study of these objects, since they are directly related to cosmic-ray origin and acceleration models. Multi-wavelength data on SNRs are sparse, and no spatially-resolved spectra are available in the $5-20$ GHz range (critical for model assessment), even for the most studied, brightest objects. Exploitation of  the high-fidelity imaging capabilities of the SRT will allow us to obtain multi-frequency, spatially-resolved information in this critical frequency range  for  complex SNRs. This will help to disentangle the different populations and different spectral behaviours of radio/gamma-ray-emitting electrons, and to obtain constraints on the high-energy emission arising from hadrons.

\subsubsection{Radio galaxies and diffuse emission in galaxy clusters}
The SRT can also make an important contribution to the
investigation of intracluster magnetic fields  (see \citealt{govoni17}). Determining the origin of
these fields and how they evolved over cosmic times, from their genesis
in the primordial Universe up to the $\mu$G levels observed in nearby 
galaxy clusters, is one of the major challenges in modern astronomy.

The most spectacular and direct evidence for the presence of relativistic
particles and magnetic fields in galaxy clusters is given by the
observation of diffuse radio sources of synchrotron radiation at the
center and in the periphery of these systems (halos and relics, e.g.,
  \citealt{ferrari08,feretti12}).
Usually, the analysis of the total intensity and polarimetric properties
of these radio sources are performed with interferometers, due to their
better spatial resolution. Nevertheless, the full extent of these radio
sources cannot be properly recovered by interferometers, especially at
frequencies $\ga$1 GHz. This issue is particularly relevant in the
context of radio spectral studies, which play a crucial role in obtaining insights 
into the mechanisms producing the halo and relic emission. 
An incomplete recovering of the flux density from 
extended radio structures could indeed lead to the derivation of incorrect spectral
properties. Sensitive single dishes providing reliable measures of large-scale, low surface brightness radio structures
are clearly valuable for this field of research.  The SRT L-band and C-band receivers, as well as 
the upcoming 7-beam S-band receiver, can all be exploited to provide such critical information on sizable samples of galaxy clusters. 

Complementary evidence of the presence of intracluster magnetic fields
is obtained through polarimetric observations of powerful and extended radio
galaxies. The presence of diffuse magnetized plasma  between the observer and a
targeted radio source changes the properties (mainly the polarization angle) of the incoming polarized emission.
From this information, the magnetic field of the intervening medium
can be inferred. Polarimetric observations of cluster radio galaxies
performed with the C-band, and the multi-feed K-band SRT receivers will 
allow us to determine the magnetic field strength and structure in galaxy 
clusters (e.g., \citealt{carilli02,murgia04,govoni04,murgia11}). 
Furthermore, by supplementing this information with SRT wide-band spectral studies, it will be possible to make a step 
forward in our understanding of the interplay between the intracluster medium
and  life-cycles of cluster radio galaxies.

\subsection{High resolution spectroscopy}
The high spectral resolution performance of
XARCOS allows us to derive precise measurements of the line-of-sight (l.o.s)
velocities of the emitting gas, which is a fundamental prerequisite for a
number of studies involving radio line observations. Among these studies, it is worth 
mentioning those aimed at placing stringent limits on fundamental constants as a function of redshift
using measurements of redshifted molecular line transitions (such as the 12 and 48~GHz methanol lines at redshift $\sim 1$,  observable at C- and K-bands respectively; 
see \citealt{bagdonaite13,kanekar15}), or at deriving black-hole masses and host-galaxy distances
through single-dish monitoring (and VLBI follow-ups) of 22~GHz water maser features in AGN
accretion disks (see, e.g., \citealt{reid13}, and references therein). Spectral
resolutions down to tens of m/s or better (at 22 GHz), are indeed crucial for the
aforementioned studies,  as they allow us to  disentangle  
narrow line features and to significantly reduce velocity measurement uncertainties.

As outlined in Sect.~\ref{sec:SRT}, in addition to providing narrow-band high spectral resolution performance in one band, XARCOS  simultaneously offers other three bands at progressively increasing bandwidth (and decreasing spectral resolution).  This makes XARCOS a versatile backend, able to address a variety of spectroscopic applications. 

\subsection{Space Applications}
The SRT will be involved in planetary radar astronomy and space missions (\citealt{tofani08}; \citealt{grueff04}), under an agreement signed by  INAF and ASI, which regulates the use of the instrument for space applications.  A detailed plan of  the activities involving the SRT is under development. In addition, the SRT is involved in a {\it Space Awareness} programme, aimed at monitoring  space debris. 

\section{Telescope Performance}
\label{sec:telperf}

In this section, we report on the results of extensive "on-sky" characterization of three main telescope specifications: pointing accuracy; primary beam response; and gain. These tests provide crucial information in view of a full assessment  of the telescope scientific capabilities, as  described in the following sections. 

\subsection{Telescope pointing}\label{sec:telpoint}

In order to test the stability of the telescope pointing, 
observing campaigns of radio sources selected from 
the Green Bank Telescope (GBT) Pointing Calibrator Catalogue\footnote{We used the new PCCALS4.7 version, kindly provided to us by J. Condon. } were performed as part of the observatory activities (see Sect.~\ref{sec:obs-tools}). 

A description of the main parameters of the Catalogue is reported in 
\cite{condon09}. From the main catalogue, we have extracted only those sources 
labeled as {\it Gold Standard} sources. These 
sources satisfy three criteria (\citealt{condon09}): i) 7mm flux densities $S \geq 0.4$ mJy; ii) accurate core positions measured by 
long-baseline interferometers; and iii) unresolved source at GBT resolution. 
The resulting catalogue of {\it Gold Standard} sources includes 570 entries. Ten 
sources routinely observed for pointing calibration purposes at the Effelsberg
100-m radio telescope and/or at the Medicina 32-m radio telescope (that were not 
present in the GBT {\it Gold Standard} catalogue) were added to the list. The final catalogue 
of putative pointing calibrators for the SRT thus lists 580 sources (see \citealt{tarchi13}).
All 580 targets and a sub-set of 260 were selected as suitable pointing calibrators for K- and C-bands, respectively. 
This selection was based on the maximum telescope beam width for which each calibrator can be used. This information is catalogued, and  accounts for possible confusion by nearby discrete sources or extended radio structure. We note that due to this selection, only 24 sources turned out to be suitable calibrator candidates for L-band observations with the SRT.

\begin{table}[t]
\caption{Statistics of the positional offset and source FWHM  for candidate pointing calibrators. }
\label{tab:cal}      
\centering          
\begin{tabular}{ c l r c}     % 5 columns 
\hline\hline
Band &    \multicolumn{1}{c}{Type}     &  \multicolumn{1}{c}{Mean} &     $N_{scans}$ \\
   &      &  \multicolumn{1}{c}{(arcsec)} &    \\
\hline    
C  &  Az offset &    $8.0\pm 0.4$ &      353\\
    & El offset &   $-6.5\pm 0.5$ &                      379\\
    & Az FWHM  &  $155.4\pm 0.7$&                       353\\
    & El FWHM  &  $157.2\pm 0.6$&                     379\\
\hline                  
K  & Az offset &  $-5.2\pm 0.4$      &  936   \\
           & El offset & $ -2.3\pm 0.3$      &  940   \\
           & Az FWHM   &  $ 52.3 \pm 0.1$       &  936   \\
           & El FWHM   &   $51.4  \pm 0.2$     &  940   \\
 \hline                  
\end{tabular}
\end{table}

The calibrators were observed with the Total Power backend 
in Azimuth/Elevation (Az/El) double cross-scans in order to determine a) the source centroid offset
with respect to the commanded position, and b) the Full Width at Half Maximum (FWHM)
of the source fitted profiles, to verify whether the targets are point-like with respect to the SRT HPBW.
The observations were performed at central observing  frequencies of  
7.24 and 21.1 GHz for C- and K-bands, respectively, with effective bandwidths of 680 MHz and 2~GHz, respectively. 
At these frequencies, the telescope beam sizes are HPBW$\sim 157$ and $\sim 52$ arcsec (see Sect.~\ref{sec:SRT}). 
The observing frequency and bandwidth at C-band were chosen so as to avoid strong RFI (see Figure~7 in \citealt{bolli15}). 

At C-band, the cross-scans were performed at a 
speed of $v_{scan}=2$ deg/min over a span of 0.4 deg, with a sampling rate of 25 Hz and an
integration time per sample of 40 ms. At K-band, the speed was $v_{scan}=1$ deg/min over a span of 0.25 deg,
with a sampling rate of 50 Hz and an integration time of 20 ms.
C-band observations were organized in seven runs spanning from the beginning of April 2014 to the 
end of August 2014.  We were able to acquire enough data for 200 out of 260 
targets, over the full 24-hour range in Right Ascension and the full 
declination range (Dec $> -40$ deg) of the catalogue.
K-band observations were carried out in five blocks:  April and August 2014, April, May and December 2015. For 79\% of the targets 
(456 out of 580),  good data were obtained for the entire RA and Dec ranges.

After an accurate flagging of the cross-scan files, the source parameters (flux density, positional offset and FWHM) were measured 
for each target and for each observing run. The average values of the source FWHM and of the positional offsets along the Elevation (El) and Azimuth (Az) axes are reported in Table~\ref{tab:cal}. The last column reports the total number of scans on which the measurements are based.
All targets were found to be very close to point-like sources, and bright enough to be 
validated as SRT pointing calibrators. The measured average FWHM of the sources are consistent with the telescope HPBW at the central observing frequency. The offsets are always found to be within the tolerance values 
($\la 10\%$ of the HPBW), demonstrating 
the stability of the SRT pointing model at least over the time spanned by the observations. We note that the offset values at K-band can be larger than the measured residuals of the pointing model (3.5 arcsec, see \citealt{bolli15}), due to the fact that various atmospheric conditions were sampled by the observations.

The main parameters of  all validated calibrators are available in  \cite{ricci16}. 

\begin{figure}
\vspace{0.5cm}
\centering
\resizebox{0.95\hsize}{!}{\includegraphics[angle=0]{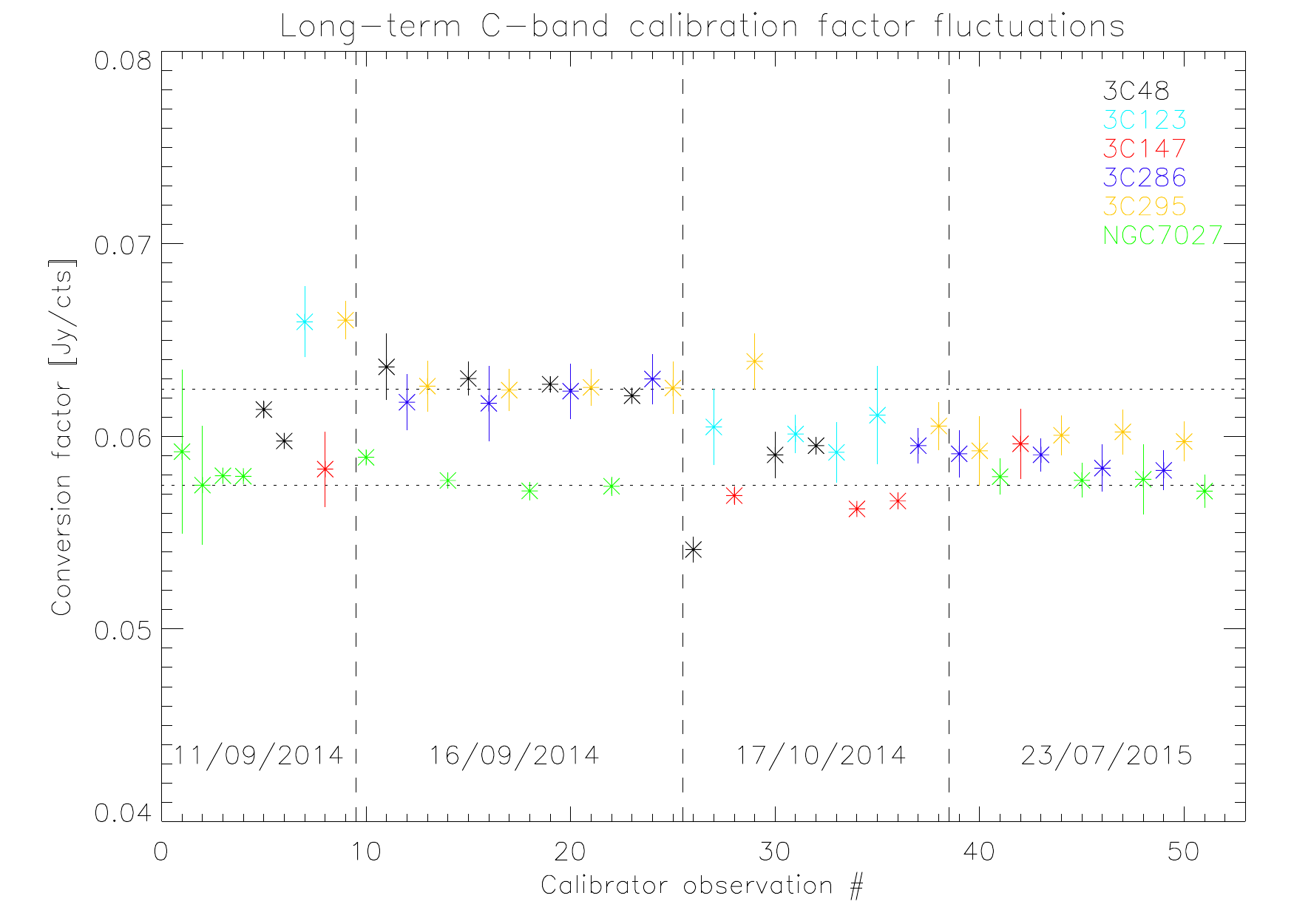}}
\caption{C-band Jansky-to-counts conversion factor as measured for different calibrators in four observing sessions (separated by dashed vertical lines), spread over a 10-month timescale. Different colors correspond to different calibrators, as labeled in the panel. The dashed horizontal lines indicate the $\pm 0.002$ rms variations around the average value of 0.06. For each observing session we report the date of the observations. }
\label{fig:gainstability}
\end{figure}

Additional calibrator observations confirmed these results; in addition, they  have allowed an estimate of the stability of the calibration conversion factor (Jy/counts) over very long timescales.
As an example, Figure~\ref{fig:gainstability} shows the C-band calibration factor fluctuations for four
observing sessions spread over a ten month period (September 2014-July
2015). Each point on the plot is related to a calibration measurement
obtained from a cross-scan (4 passes).
The average Jy/cts factor at 7.2 GHz (bandwidth 680 MHz) is $0.060\pm 0.002$, meaning long-term fluctuations 
<5\%. The reported calibration factor fluctuations mostly depend on weather conditions (opacity fluctuations), RFI affecting
cross-scan measurements, calibrator flux density uncertainties, and actual
instrumentation gain fluctuations.
In the case of optimal weather conditions, and selecting very stable calibrators
(like e.g. 3C286) observed at $>20$\deg~elevations, the calibration factor
fluctuations are below 2\%.

 \begin{figure*}
%\centering
%\vspace{-1.5cm}
\resizebox{12cm}{!}{\includegraphics[angle=0,clip]{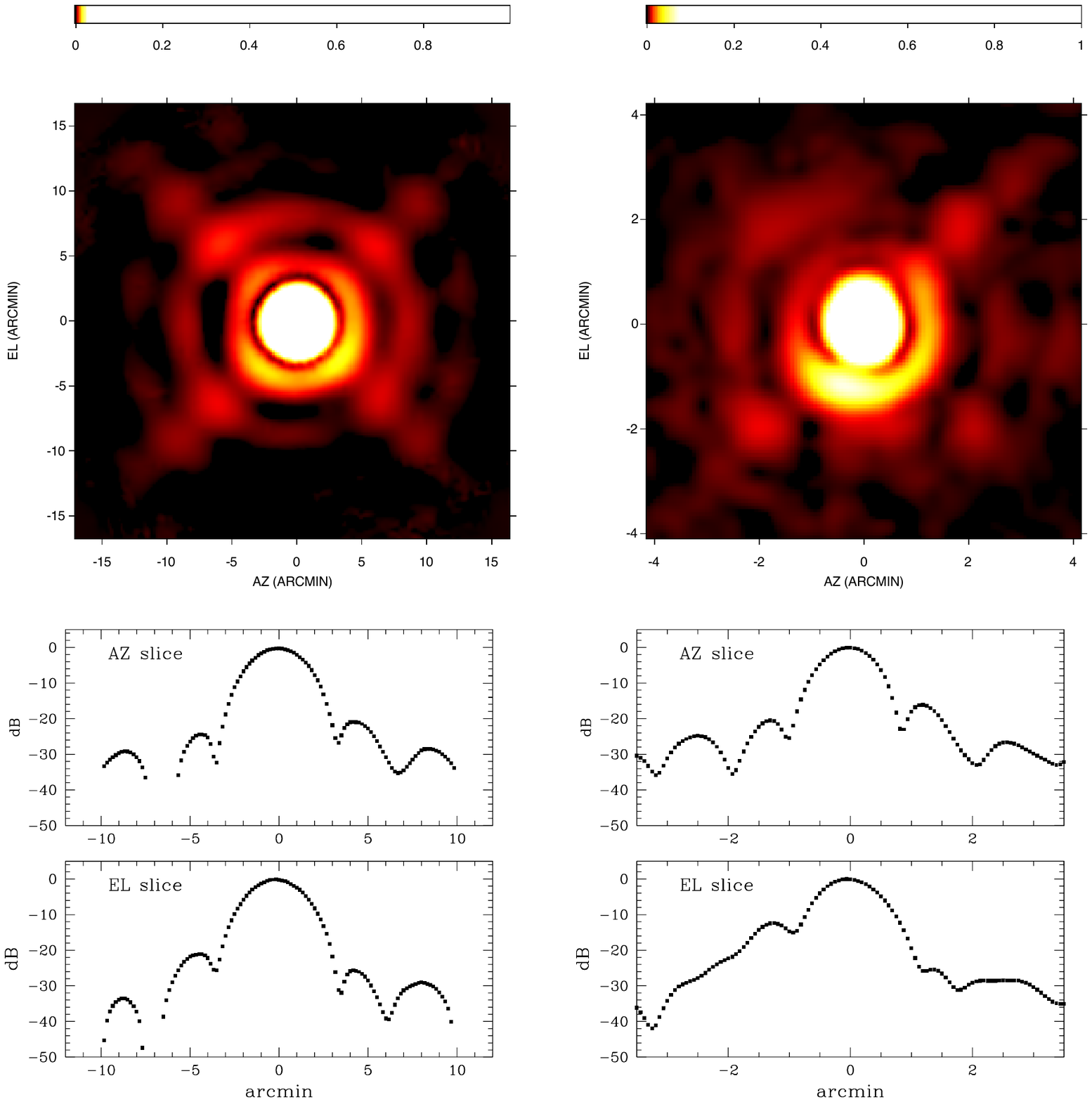}}
\hfill
\begin{minipage}[t]{55mm}
\vspace{-11cm}
\caption{SRT beam patterns at C--band  (7.24 GHz, left-column panels) and K--band (right-column panels) respectively. For K--band we show only the 25.54 GHz measurement of the central beam. {\it Top:} elevation-averaged beam patterns (normalized to a peak value of 1) obtained by
stacking together several OTF scans of bright point sources taken at different elevations. The color map has been intentionally saturated to highlight the low intensity features of the second and third lobes. {\it Bottom}: beam cross-sections along the Azimuth and Elevation axes (in dB).}
\label{fig:beam}
\end{minipage}
\vspace{-5.5cm}
\end{figure*}
        
\subsubsection{Tracking and Scanning Accuracy}
For mapping experiments like those reported in Sect.~\ref{sec:imaging}, it is important to verify whether the pointing accuracy is affected by the scanning speed.
The tracking and scanning accuracy of the telescope was verified  through OTF scans. Acquisitions consisted in Az/El cross-scans performed at different scanning speeds, at various Elevation positions. 
SRT users are requested to indicate a `sky speed' (i.e. the actual scanning speed measured `on sky') in the observing schedules. For Elevation sub-scans, such speed corresponds to the speed around the Elevation axis. When scanning along Azimuth, on the other hand, the sky speed translates into Azimuth `axis speed' according to:
\begin{equation}
{\rm AzAxisSpeed} = {\rm SkySpeed}/{\rm cos(El)}.
\end{equation}
 When performing each sub-scan, the antenna tracks the source in both Azimuth and Elevation, while running along a pre-defined path on the `scanning axis'. Incremental offsets are applied to the scanning axis motion, varying with a constant step within the defined span, with respect to the instantaneous source position. Our data can thus indicate both the tracking (i.e. along the non-scanning axis) and the scanning accuracies. 

This test was performed at K-band only, i.e. where the pointing accuracy tolerance ($\la 5$ arcsec) is the most stringent. In particular, we chose to observe in the $24.00-24.68$ GHz sub-band, where the SRT beam-size (HPBW$=46^{\prime\prime}$) is smaller and both the RFI and opacity contributions are less prominent (with respect to lower frequencies within the K-band receiver RF range). 

The pointing accuracy along the tracking axis turned out to be stable at the sub-arcsec level, for any scanning speed. The positional offsets along the scanning axis is in line with the measured pointing offsets listed in Tab.~\ref{tab:cal}, and proved to remain tolerable for 'axis speeds' up to 20$^{\circ}$/min. Deviations from this level of performance, showing pointing offsets up to $15^{\prime\prime}$, have been detected in a limited number of sessions (mainly for the Elevation scans), regardless of the scanning speed. This phenomenon is likely related to thermal deformations of the telescope structure, indicating that the implementation of metrology-based corrections might be required, in the case of significant thermal variations, even when operating at K-band. 

\begin{table}[t]
\caption{Observational details of the SRT beam characterization campaigns at C-- and K--bands using the TP backend.  }
\label{tab:beam}      
\centering  
\footnotesize        
\begin{tabular}{ c l r  c c c}     % 5 columns 
\hline\hline
Band &   Target     &  $\nu_{obs}$   &  FoV & $v_{scan}$ & N$_{scans}$\\
         &                     &  (GHz)          &  (arcmin$^2$) & ($^{\prime}/s$)    \\
\hline    
C($^\dag$)   &  3C~147 &  7.24  & $18 \times 18$ & 3 & 288\\
                    &  3C~273 &  7.24  & $18\times 18$ & 3  & 86\\
\hline                  
K$(^\bot)$  & 3C~84 &  18.34   & $12\times 12$  & 6  & 20\\
    & 3C~84  &  22.04   & $12\times 12$  & 6  & 20\\
    & 3C~84  &  23.74   & $12\times 12$  & 6  & 20\\
    & 3C~84 &  25.54   & $12\times 12$  & 6  & 20\\
 \hline                  
\end{tabular}
\tablefoot{
\tablefoottext{$\dag$}{Observations undertaken in the period April 2014 -- April 2015.}
\tablefoottext{$\bot$}{Observations undertaken in October 2015.}
}
\end{table}

\begin{figure*}
\centering
\vspace{-6cm}
\resizebox{12cm}{!}{\includegraphics[angle=0]{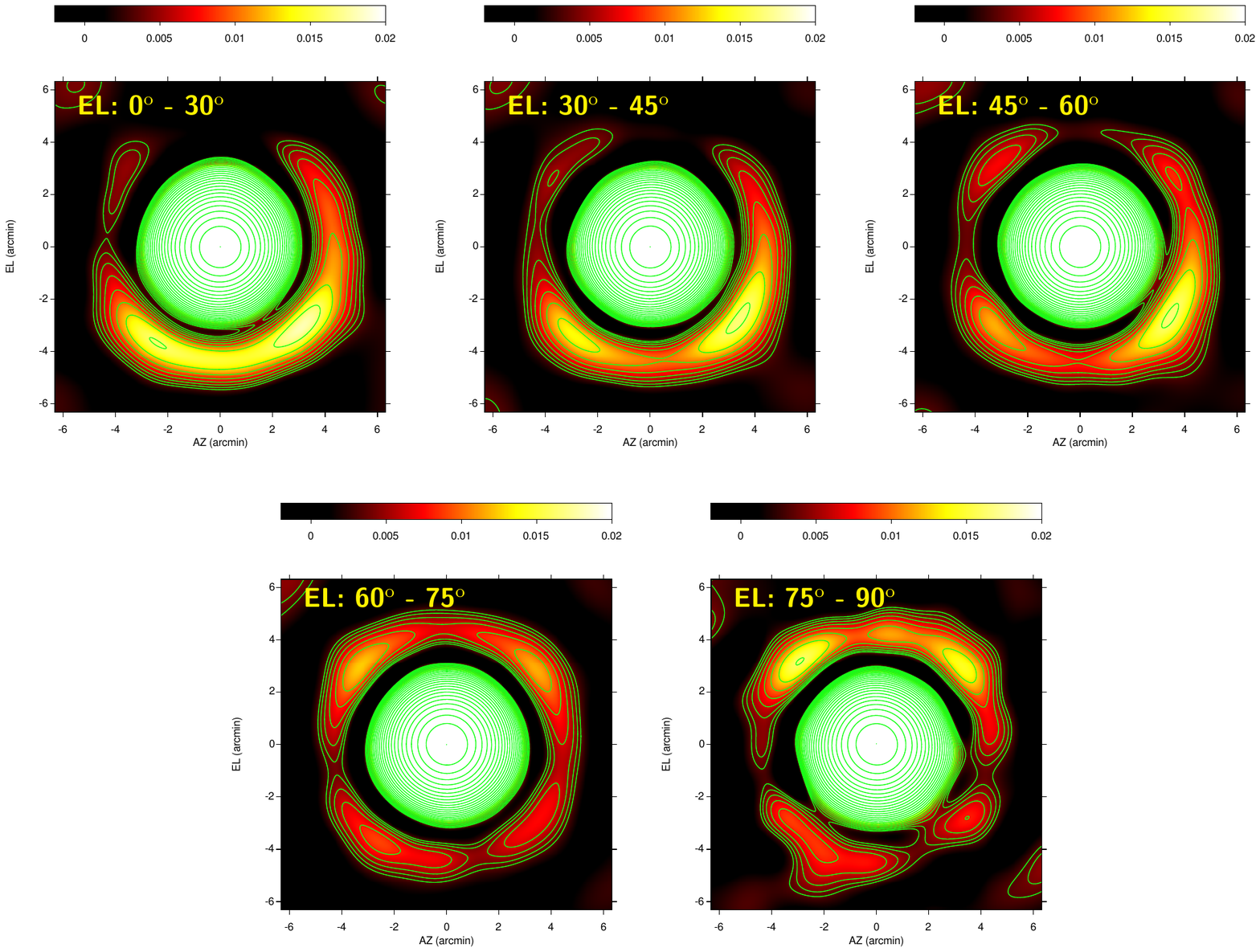}}
\hfill
\begin{minipage}[t]{55mm}
\vspace{-4.7cm}
\caption{Variation of the SRT beam pattern at 7.24 GHz as a function of elevation. Each image shows a field-of-view of $6^{\prime}\times 6^{\prime}$ centered on the beam pattern peak (normalized to 1). Contour levels start at -25 dB and increase by 1 dB.}
\label{fig:beamEl}
\end{minipage}
\vspace{-1.5cm}
\end{figure*}

\subsection{Detailed Characterization of the Beam Pattern}\label{sec:beam}

The SRT beam shape was characterized as part of the technical commissioning activities (see \citealt{bolli15}), where residual deformations of the main lobe were measured as a function of elevation, once the SRT optical alignment was optimized and the active surface was implemented. Here we provide a finer characterization of the SRT beam shape (and its possible variations as a function of telescope elevation), by extending the analysis from the main lobe only to the third lobe (or second side-lobe). This finer beam characterization is particularly important for high dynamic range imaging in the presence of bright sources (see Sect.~\ref{sec:HDR-3C147} for a demonstration). In this case, beam deconvolution is required, as the lateral lobes of the beam, if poorly known, can limit the final sensitivity of the image.

In this section, we report on the results obtained at C-band and at K-band, where a full characterization of each of the 7 beams was obtained. 
The beam shapes at C- and K- bands  were measured through a campaign of OTF scans in all possible frames (equatorial, horizontal, and Galactic), collecting a large number  of maps along RA, Dec, Az, El, GLON (Galactic Longitude), and GLAT (Galactic latitude) directions.  We sampled along rotated frames to reduce the artifacts related to the odd OTF sampling. 
The data were acquired using the Total Power backend. At K--band, we observed at four different frequencies, distributed in the 8~GHz frequency range covered by the K-band receiver ($18-26$ GHz). The seven beams of the K-band feed array were characterized simultaneously. 

Table~\ref{tab:beam} reports the observational details of the C-- and K--band campaigns. For each target, we list the central frequency of the observations ($\nu_{obs}$);  the field of view (FoV), i.e. the region around the target that was imaged; the mapping speed ($v_{scan}$); and the number of scans performed (half for left and half for right polarizations). The sampling interval and integration time were always set to 10 ms. The instantaneous bandwidth was always set to 680 MHz.

For data calibration and analysis, we used the SCUBE data reduction software (\citealt{murgia16}, see Sect.~\ref{sec:tools3} for a brief description). The beam pattern was obtained through iterative wavelet modeling: an increasingly refined baseline subtraction and a `self-calibration' were applied to remove 
the residual elevation gain variations and pointing offsets from the individual OTF scans.

The self-calibrated OTF scans were combined to derive a detailed model of the SRT beam pattern at C-- and K--bands down to the third lobe (see Fig.~\ref{fig:beam} left--column  and right--column panels respectively).   
The elevation-averaged beam patterns (in Jy/beam) are shown in the top panels. It is evident that we successfully detected the third lobe of the SRT beam pattern. The bottom panels show two perpendicular beam cuts intercepting the peak. One is directed along the Azimuth axis (upper panel) and the other along the Elevation axis (lower panel).  

The expected size of the beam pattern main lobe at 7.24 GHz is HPBW=2.6 arcmin. The inner 4$^{\prime}\times 4^{\prime}$ of the main lobe of the beam pattern at this frequency was measured by fitting the target source with a 2D elliptical Gaussian with four free parameters: peak; minimum and maximum full-width half-maximum sizes (FWHM$_{\rm min}$ and FWHM$_{\rm max}$); and position angle (PA). We found FWHM$_{\rm min}=157^{\prime\prime}$ and FWHM$_{\rm max}=158^{\prime\prime}$, with PA $\simeq 0^{\circ}$, confirming that the SRT main lobe at 7.24 GHz can be considered to be circularly symmetric with FWHM$=2.62^{\prime}$.
The secondary lobe intensity averaged over an annulus of 4.5$^{\prime}$ radius and 3$^{\prime}$ width is  -24 dB. The secondary lobe is clearly asymmetric, being brighter below the main lobe. We anticipate that this asymmetry is elevation dependent, as further discussed below.
The average intensity of the third lobe over an annulus of 9$^{\prime}$ in radius and 3$^{\prime}$ in width is  -33 dB. The most striking features associated with the third lobe are the four spikes seen at the tips of a cross tilted by 45$^{\circ}$. These originate from the blockage caused by X-shaped struts sustaining the secondary mirror in the path of radiation between the source and the primary mirror. 

\begin{figure*}
\vspace{-15.5cm}
\centering
\resizebox{0.9\hsize}{!}{\includegraphics[angle=0]{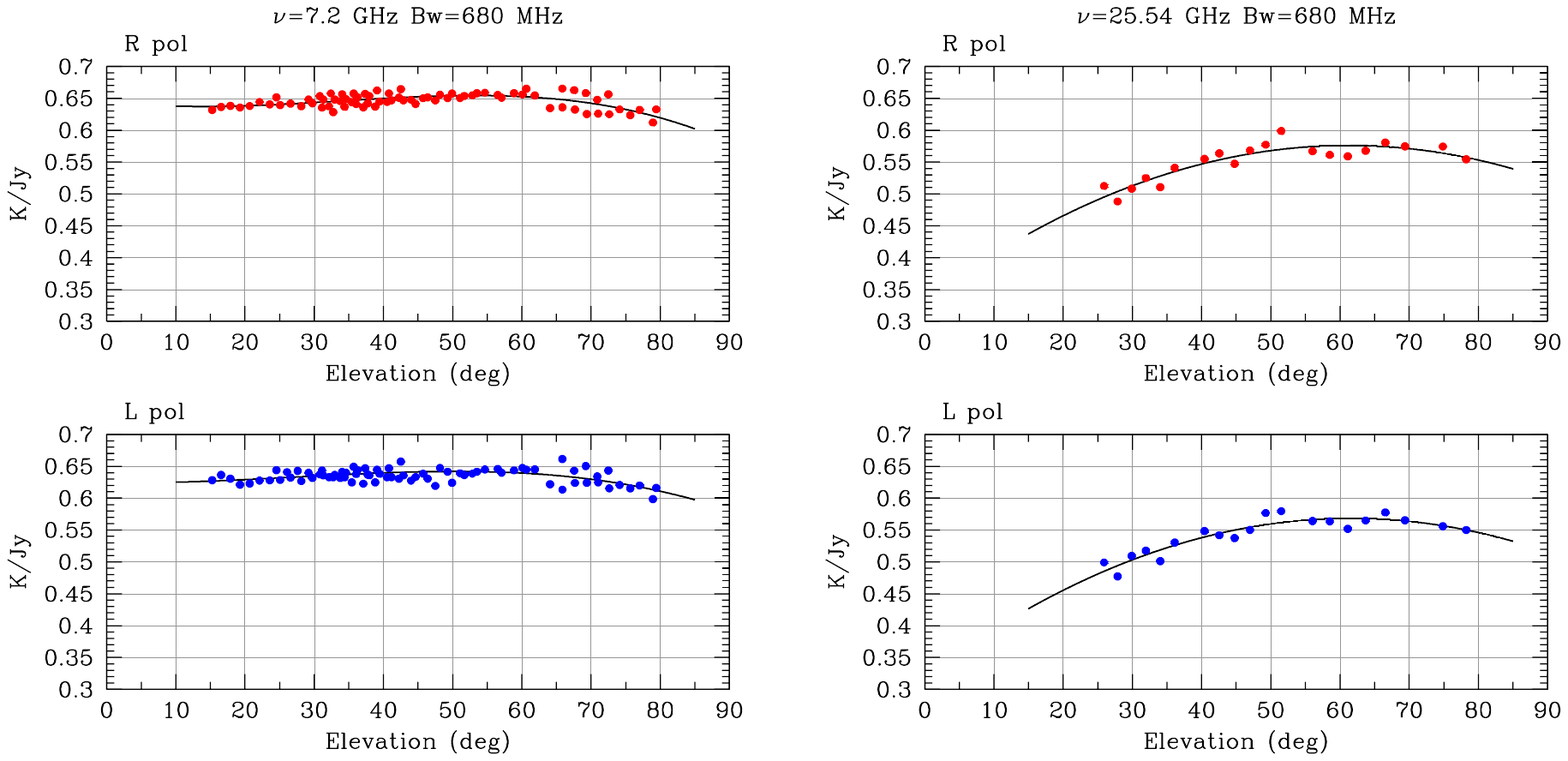}}
\caption{Gain (in K/Jy) as a function of elevation measured through OTF scans at 7.24 GHz (left panels) and at 25.54 GHz (right panels), with a bandwidth of 680 MHz. The right and left polarizations are shown separately (top and bottom panels respectively). The solid lines represent fourth order polynomial fits. The 25.54 GHz gain curve refers to the central feed of the K-band receiver.}
\label{fig:gaincurve}
\end{figure*}

To study the variation of the beam shape with elevation, we grouped the OTF scans at C--band  into five bins of 15$^{\circ}$ in width. A zoom of the inner $6^{\prime}\times 6^{\prime}$ showing variations of the second lobe asymmetry with a slightly finer elevation binning is presented in Fig.~\ref{fig:beamEl}.
It is clear that the beam pattern varies significantly with elevation. The most important effect is the behavior of the secondary lobe structure at high elevations. In particular, we observe a flip of the asymmetry around an elevation of 60$^{\circ}$. At elevations higher than 60$^{\circ}$, the brightest part of the secondary lobe is observed above the main lobe, while it is observed below it at mid and low elevations.  The same behaviour is observed for the K-band beam (not shown).

The expected size (HPBW) of the beam pattern's main lobe at K-band varies from $\sim 1$ arcmin (at 18 GHz) to $\sim 0.7$ arcmin (at 26 GHz). Here we report on the results obtained at 25.54 GHz for the central feed (shown in Fig.~\ref{fig:beam}, right panels). The results obtained for the other three frequencies, as well as for the other feeds, are qualitatively similar. 
By fitting the target source with a 2D elliptical Gaussian with four free parameters, as done at C-band,  we found FWHM$_{\rm min}=43^{\prime\prime}$ and FWHM$_{\rm max}=48^{\prime\prime}$, with PA $\simeq 0^{\circ}$. This clearly indicates that the SRT's main lobe at K-band is no longer circularly symmetric, but rather slightly elliptical.  The asymmetry of the secondary lobe is even more pronounced, with a very bright peak below the main lobe (-12 dB, see Elevation cut). 
This asymmetry does not satisfy the SRT design specifications (secondary lobe below -20 dB).

The current LUT was built through photogrammetry measurements in six elevation positions. Our tests seem to suggest that denser elevation-dependent measurements are needed to correct the residual misalignments of the telescope optics, which affect the K-band beam pattern.  In addition, the LUT of the secondary mirror was built during the first SRT optical alignment (see \citealt{bolli15}), and was mainly aimed at correcting the much larger primary mirror deformations.  A laser tracker measurement campaign is currently ongoing to map the position of the secondary mirror as the telescope varies in its elevation angular range with and without using the LUT. Thanks to the sub-millimetre accuracy of the laser tracker\footnote{We used a Leica Absolute Tracker AT402 model.}, we can finely measure the translations and the rotations of the secondary mirror at several elevation angles of the telescope, and refine the LUT, mainly at low (but even at high) elevation angles where some residual misalignments are still present. 

We point out, however, that even in the present situation, the SRT offers excellent imaging capabilities, in particular when combined with deconvolution techniques. This will be illustrated in Sect.~\ref{sec:imaging}.  

\subsection{Fine Characterization of the Gain Curve}\label{sec:gaincurve}

An important by-product of the in-depth characterization of the primary beam at C- and K-bands (see Sect.~\ref{sec:beam}) is a very reliable measurement of the gain curves at fine steps in elevation. Indeed, the best fit peak amplitude obtained through self-calibration, divided by the target flux density, directly yields the antenna gain in K/Jy. 
In Figure~\ref{fig:gaincurve} we present the gain curves measured at 7.24 GHz (left panels) and at 25.54 GHz (right panels), for the right (top) and left (bottom) polarizations separately.   

At C-band, the gain curves were derived using the observations of the calibrator 3C~147 (see Sect.~\ref{sec:beam}) and by assuming a flux density of 5.43 Jy at 7.24 GHz for the target (\citealt{baars77} scale). 
The measured median gains are  0.63 and 0.62 K/Jy for right  and left  polarizations respectively, with a scatter of about $8-9\%$. These values are slightly higher than those measured during the SRT technical commissioning activities (see \citealt{bolli15} and our Table~\ref{tab:receivers}) and are closer to expected (see Table~2.2 of \citealt{bolli15}). 
The gain curves are extremely flat, indicating that the active surface is performing very well. 

The gain curves measured at 25.54~GHz for the central feed of the K-band receiver were obtained from the observations of the bright radio source 3C~84 (see Sect.~\ref{sec:beam}). We corrected the OTF scans for the atmosphere opacity and we used the primary calibrator 3C~295 to derive the  3C~84 flux density (26.83 Jy at 25.54~GHz; \citealt{perley13} scale). 
Figure~\ref{fig:gaincurve} (right panels) clearly shows a decrease in efficiency below 45$^{\circ}$ of elevation, while  the gain curve is flat at higher elevations. This confirms what was found during the commissioning activities (\citealt{bolli15}), and indicates that the modeling of the gravity deformations at low elevations can be improved (see Sect.~\ref{sec:beam}). We note, however, that we measure a maximum gain of $\sim 0.58$ K/Jy for both polarizations. This value is lower than the one measured during the commissioning at a frequency of 22.35 GHz (see \citealt{bolli15} and our Table~\ref{tab:receivers}). However it is still in line with the expectations for the K-band receiver (0.6-0.65 K/Jy, see Table~2.2 of \citealt{bolli15}).  

It is worth noting that the gain curves presented here are obtained through OTF imaging,  as discussed in Sect.~\ref{sec:beam}.  These curves are fully consistent with others obtained through standard gain curve procedures based on cross-scans. This novel technique based on imaging, is particularly efficient for multi-feed receivers, since it allows us to obtain the gain curves for all beams simultaneously. The measurements obtained for the lateral feeds of the K-band receiver are very similar to those for the central feed and are not shown.

The SRT gain curves  are regularly updated on the SRT website. 

\begin{figure}
\centering
\resizebox{0.9\hsize}{!}{\includegraphics[angle=-90]{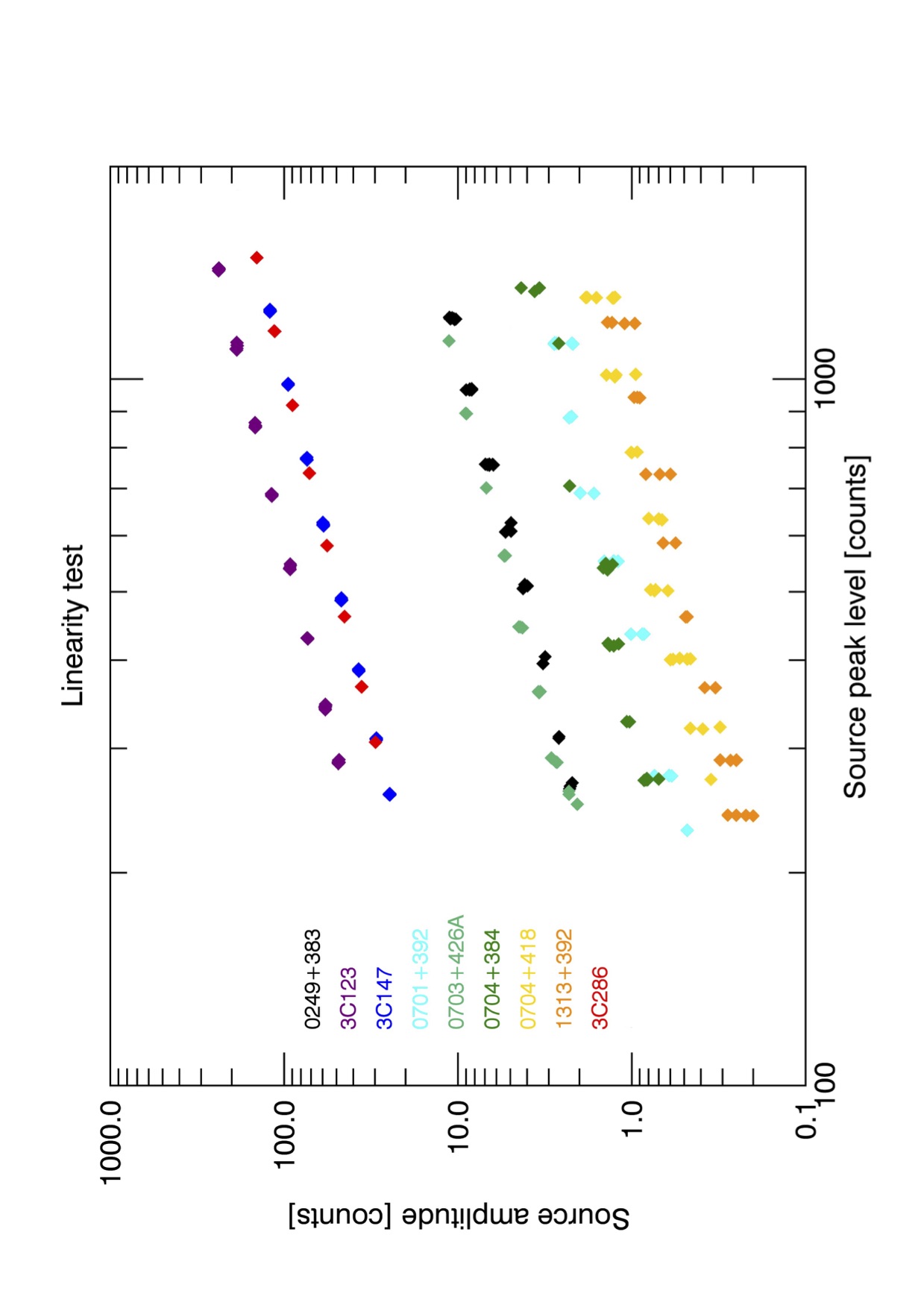}}
\resizebox{0.9\hsize}{!}{\includegraphics[angle=-90]{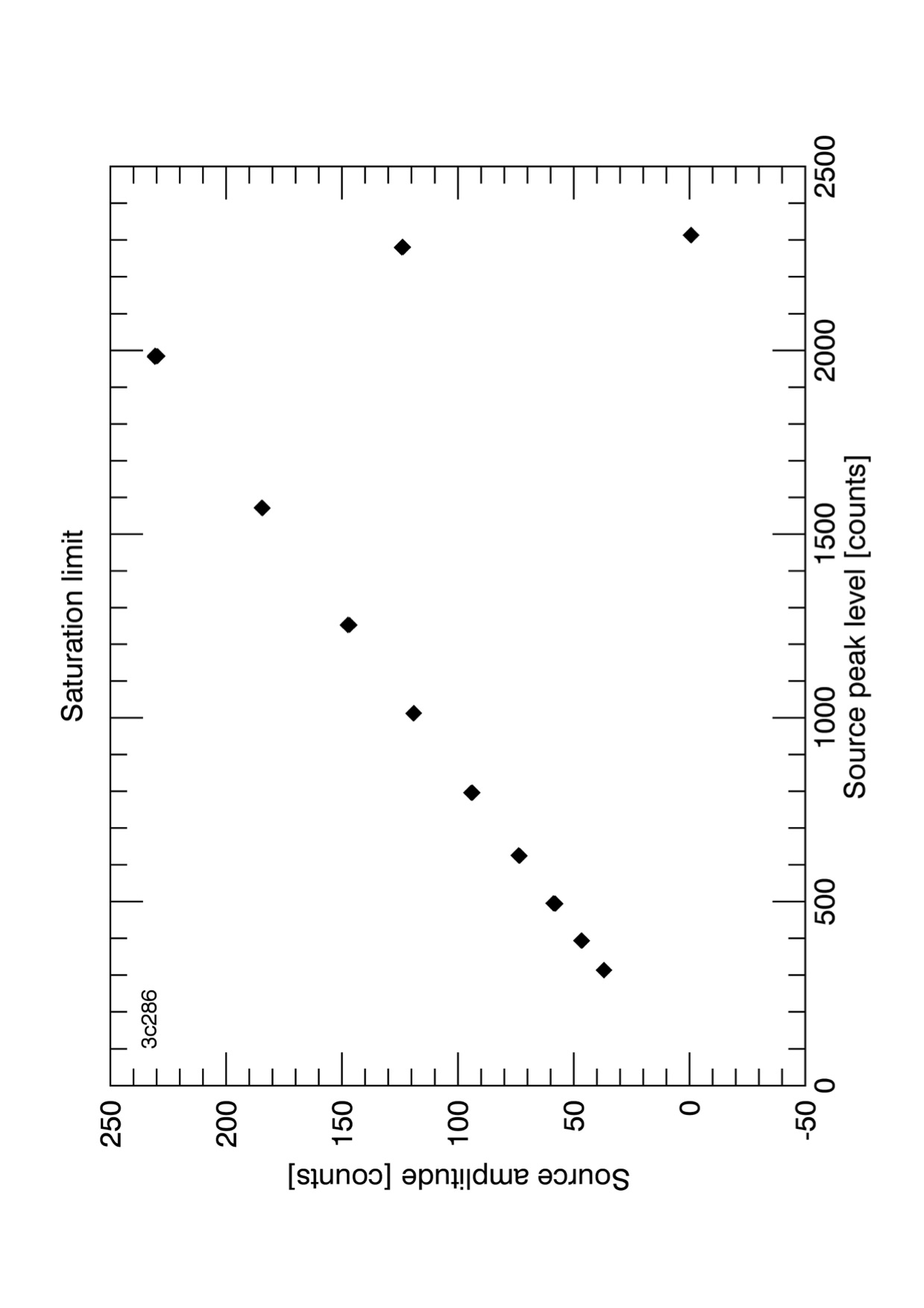}}
\caption{Source (baseline-subtracted) amplitude  vs peak (source + baseline)  level, both expressed in raw counts. Increasing raw counts on both axes reflect decreasing attenuation. 
{\it Top panel:} Test performed with 250 MHz bandwidth at a central observing frequency of 7.25 GHz. Different sources are indicated by different colors, as labeled in the panel. {\it Bottom panel:} Test performed with 1200 MHz bandwidth at a central observing frequency of 6.8 GHz. Here only the bright source 3C286 (5.87 Jy at 6.8 GHz, \citealt{perley13}) is shown, to highlight the levels of raw counts where saturation is reached.}
\label{fig:GZT05}
\end{figure}

\section{Radio-Continuum Observations with the Total Power backend}\label{sec:tpperf}
In the following, we report the on-sky characterization of the TP backend. The tests discussed here  provide a useful reference for future telescope users. In addition, a good characterization of the TP is a prerequisite for efficiently conducting high-performance imaging experiments, such as those discussed in Sect.~\ref{sec:imaging}. 

\subsection{Backend linearity}
\label{sec:linearity}
The response of the TP backend was tested through the execution of cross-scan observations of a selection of well-known sources from the B3-VLA catalogue (\citealt{vigotti99}), with the addition of a number of other  widely-used calibrators. This test, which is band-independent, was carried out at C-band. The selected sources range in flux density from a few tens of mJy to a few Jy (flux densities measured at 4.85 GHz).  The observations were performed for varying bandwidths and attenuation settings, in order to span a wide range of `raw counts' signal levels and probe the backend linearity range. The results are summarized in Fig.~\ref{fig:GZT05}.  The top panel of Figure~\ref{fig:GZT05} shows the results obtained for different sources  by setting a bandwidth of 250 MHz.  It is clear that the backend response is linear over a quite wide range of raw counts, from 
$\sim 200$ to $\sim 2000$ counts (x-axis). With a narrow band of 250 MHz, saturation is never reached, even with no attenuation ($=0$ dB). We note that for sources with flux density $<100$ mJy, saturation cannot be reached, regardless of the bandwidth or the attenuation setting. The bottom panel of Figure~\ref{fig:GZT05} shows the result for the bright source 3C286 only (5.87 Jy at 6.8 GHz, \citealt{perley13}), when setting a larger bandwidth of 1200 MHz. This plot clearly shows  that saturation is reached around 2000 raw counts. Since the backend linearity is well preserved at low baseline levels, it is recommended to work at levels of a few hundred counts (e.g. $\sim 400-500$ counts), so as to fully exploit the available dynamic range. 

\begin{figure}
\centering
\vspace{0.2cm}
\resizebox{0.9\hsize}{!}{\includegraphics[angle=0]{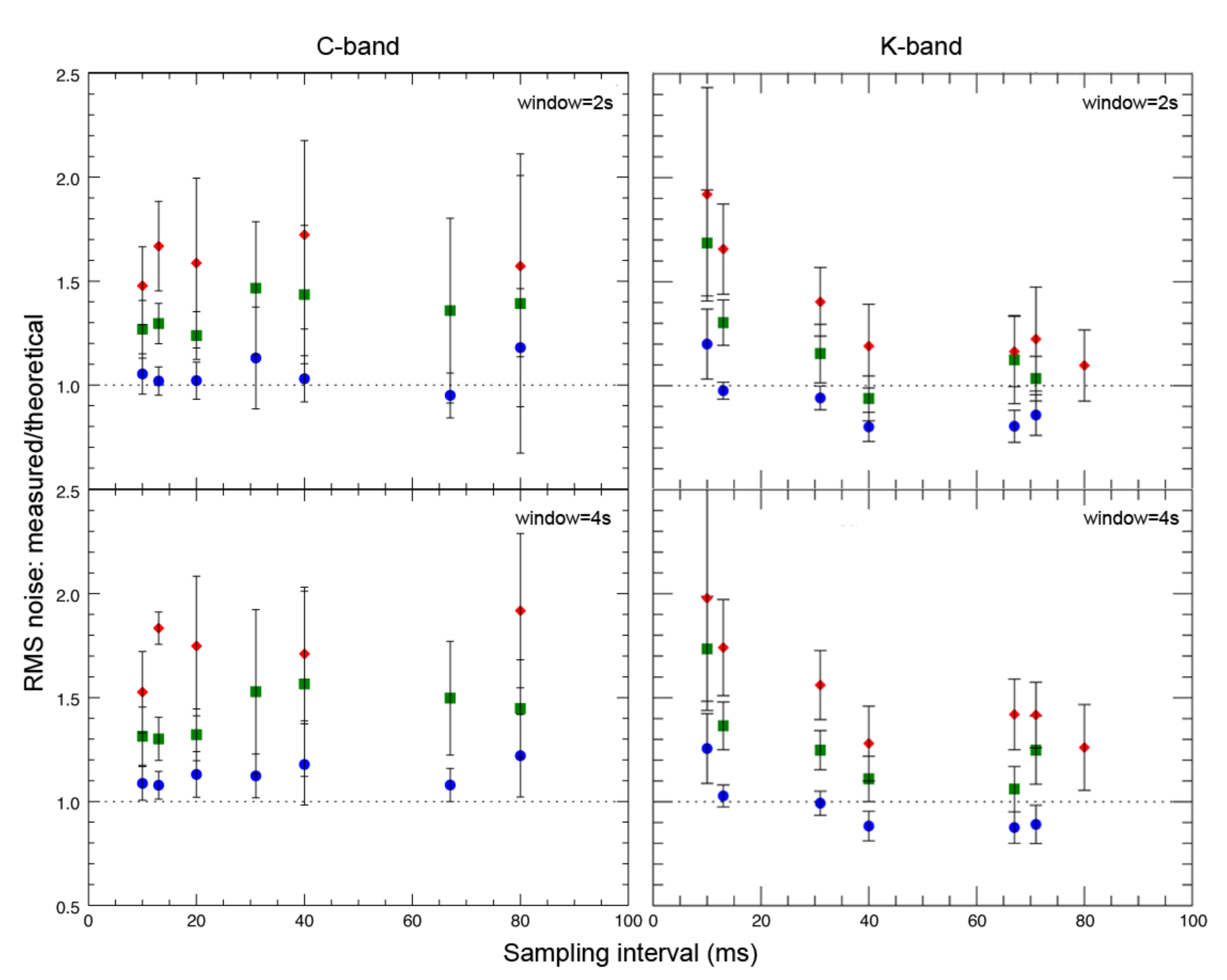}}
\vspace{0.3cm}
\caption{Ratio of measured over expected noise as a function of sampling interval for different bandwidths (blue circles=250~MHz, green squares=680~MHz, red diamonds=1200~MHz). The noise RMS is measured (and estimated) over different time windows. Here we show the results for 2~s (top) and 4~s (bottom) time windows.  The experiment was performed at both C- (left panels) and K-band (right panels). }
\label{fig:GZT01C}
\end{figure}

\subsection{Band-limited noise and confusion limit}
\label{sec:noise}
This experiment focused on the measurement of the radio continuum noise, which was recorded by selecting different bandwidths and sampling intervals. Ideally, as long as the post-detection integration time is short enough that $1/f$ noise effects are not dominant, the measured noise is supposed to be band-limited, i.e. it should coincide with a white noise whose RMS value ($\sigma$) - given a certain system temperature - depends only on the selected bandwidth and on the integration time, according to the radiometer equation:
\begin{equation}\label{eq:noise}
\sigma = \frac{T_{sys}/G}{\sqrt{BW t}},  
\end{equation}
where the antenna gain is given in K/Jy,  the system temperature in Kelvin, the bandwidth in Hertz, and  the integration time, $t$, in seconds.
The aim of this test  was to verify whether - and for which bandwidths - the thermal noise, estimated through the radiometer equation, could be reached.  This test is particularly relevant for the TP, as analog backends do not allow an efficient removal of RFI-affected data, and this can easily result in increased noise levels, especially when large bandwidths are used.

The experiment was performed at both C- and K-bands. The observing frequencies were chosen to minimize the presence of RFI  (see \citealt{bolli15}). At C-band, we set the low end of the bandwidth at 7~GHz. At K-band, we set the low end of the bandwidth at 24 GHz. We acquired data employing different setups: we varied the bandwidth from 250 to 1200~MHz, while data sampling ranged from 10 to 80~ms.  The subsequent noise RMS measurements were performed over different time windows in order to reveal possible effects and instabilities at different time scales. Figure \ref{fig:GZT01C} shows the results for C-band (left panels) and K-band (right panels), in terms of the measured observed-to-expected RMS ratio,  for either 2~s (top) or 4~s (bottom) time windows. 
All plots clearly show a band-related increase of the noise ratio, with the measured noise being very close to expected for the smallest bandwidth (250 MHz), and becoming a factor of $1.5-2$ larger when increasing the bandwidth to 1200 MHz. This is probably due to low-level RFI, whose impact on continuum observations becomes larger for wider bandwidths, as more polluting signals are likely to be gathered in a wider frequency range. This increase is more pronounced at C-band, since it is more severely affected by RFI. The noise ratio also increases with the time window. This is probably due to instabilities and gain drifts becoming visible in the 4~s time frame, leading to a measured signal which is not purely white noise anymore. These measurements allowed us to confirm that band-limited noise can be reached by the TP, at least for narrow bandwidths. Moreover, the effects due to RFI can be kept under control by properly setting the observing frequency. 

\begin{figure}
\centering
\resizebox{1.1\hsize}{!}{\includegraphics[angle=90]{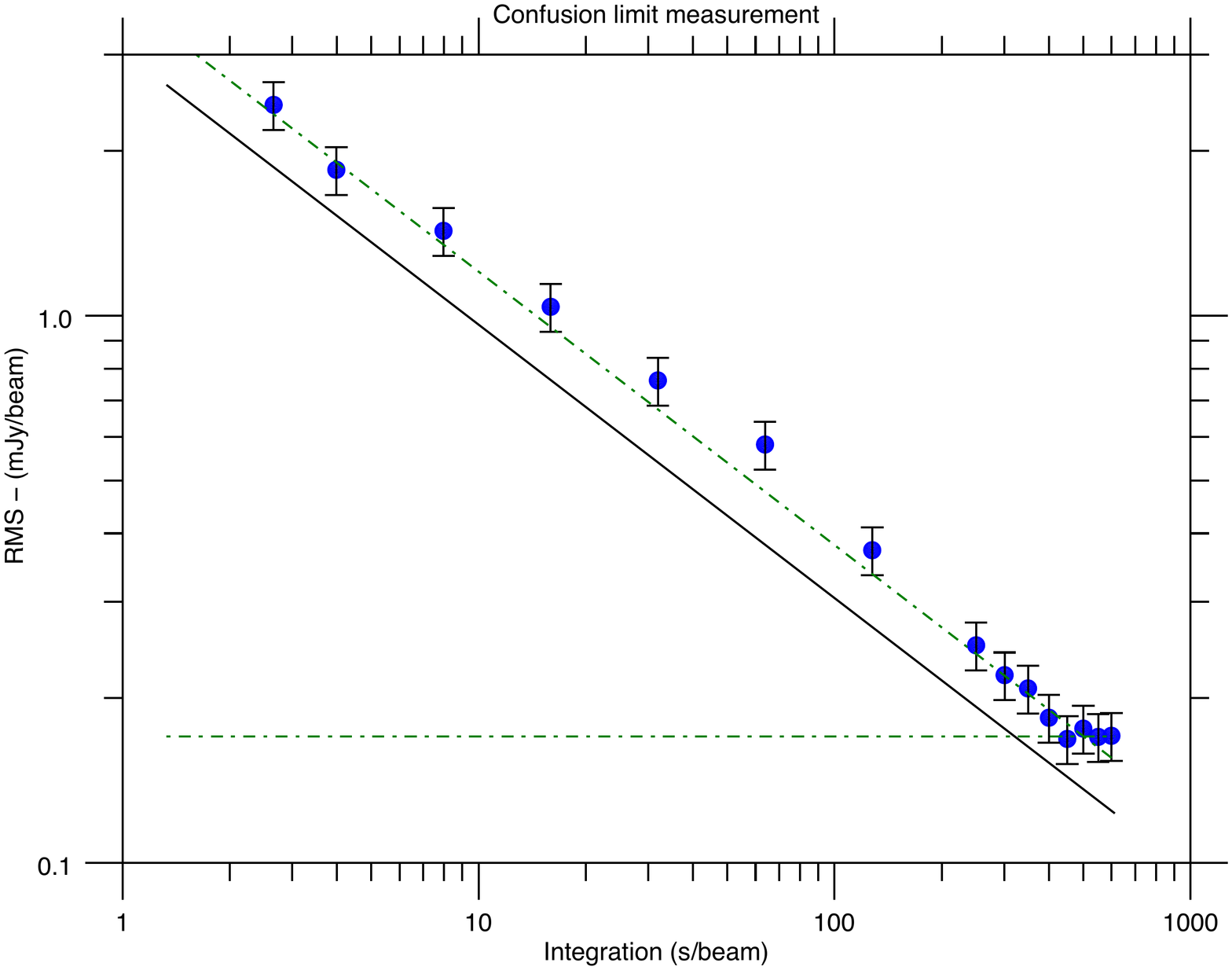}}
\caption{RMS noise as a function of integration time. The oblique solid line shows the expected trend according to the radiometer equation. The parallel dot-dashed line corresponds to the theoretical noise increased by 25\%. The horizontal dot-dashed line shows the measured confusion limit. }
\label{fig:GZT06C}
\end{figure}

Another important parameter that must be estimated - when planning observations - is the integration time needed to reach the desired sensitivity. A dedicated test was carried out in order to verify whether the noise decreased with integration time ($t$) following the expected $1/\sqrt{t}$ law described by the radiometer equation (see Eq.~\ref{eq:noise}). The experiment was performed at C-band, where the system temperature is very stable over a wide range of Elevation positions (20-80$^{\circ}$), and where the gain curve is almost flat (see \citealt{bolli15}). This means that variations of the telescope performance parameters with Elevation did not affect our measurements. 
The observations were carried out at a central observing frequency of 7.5~GHz with a bandwidth of 250 MHz. They consisted of repeated back-forth OTF scans along RA, acquired over a selected `empty sky' strip. This was chosen within a deeply-mapped area of the 10C survey (\citealt{ami11}); more specifically, the strip was centered on RA=00h27m00ss, Dec=+31$^{\circ}$35'00" and it covered an RA span of 0.5$^{\circ}$. No sources with a flux density greater than 0.1 mJy at 10 GHz are known to exist in this strip. The scans were inspected, flagged, and iteratively summed, to measure the RMS noise for increasing integration times. The results are shown in Fig.~\ref{fig:GZT06C}. Raw data were calibrated employing the frontend noise diode. Antenna temperature values were then converted into flux density by applying the gain curve. We find that the RMS noise decreases with integration time as predicted by the radiometer equation, all the way down to the so-called confusion limit, i.e the impassable noise plateau produced by the presence of background faint unresolved sources falling in the telescope main beam.  However, the RMS noise that is actually measured is $\sim 25\%$ higher than the theoretical value;  this discrepancy can be ascribed to the overall instabilities arising from long exposure times (as described above). The higher noise level, of course, implies that the confusion limit is reached in a longer-than-predicted exposure time, amounting to $\sim 430$ s/beam. 

The SRT ETC  (see Sect.~\ref{sec:tools1}) provides an estimate of the confusion noise at the center of the observing band. This estimate is based on extrapolations of known source counts to different frequencies and/or lower flux densities (for details see \citealt{zanichelli15}). Using the algorithm implemented in the ETC, we get  $\sim 0.19$ mJy at 7.5 GHz, our observing frequency. The actually measured confusion noise is  $0.17 \pm 0.02$ mJy/beam, so it is consistent with the predicted value. We can therefore conclude that the ETC provides a reliable indication of the telescope confusion limit. 
 
\section{Imaging Capabilities of the SRT}\label{sec:imaging}

In this section, we seek to demonstrate the radio-continuum imaging capabilities of the SRT.  
Such capabilities can be inferred through the assessment of the so-called {\it dynamic range} and  {\it image fidelity}. The dynamic range  (defined as the peak-to-noise rms ratio) identifies the ability to reach the thermal noise and/or to image faint or low-surface brightness in the presence of very strong sources. The image fidelity  is a measure of the reliability of the  image (in terms of surface brightness, size, and morphology), when mapping extended sources. In the following, we discuss the SRT capabilities for both the aforementioned parameters, for radio continuum observations with the TP backend, focussing on observations at C-- and K--bands.  As mentioned in Sect.~\ref{sec:SRT}, the use of analog backends, like the  TP,  at L-- and P-- bands, is not recommended due to severe RFI pollution. 

The results illustrated in this section demonstrate not only the imaging capabilities of the SRT and of its receiving/acquisition systems, but also the role that can be played by innovative ad-hoc imaging techniques, based on OTF scans and state-of-the-art data analysis software (like e.g. SCUBE and SDI, described in Sect.~\ref{sec:tools3}).

\begin{figure*}
\centering
\resizebox{0.9\hsize}{!}{\includegraphics[angle=0]{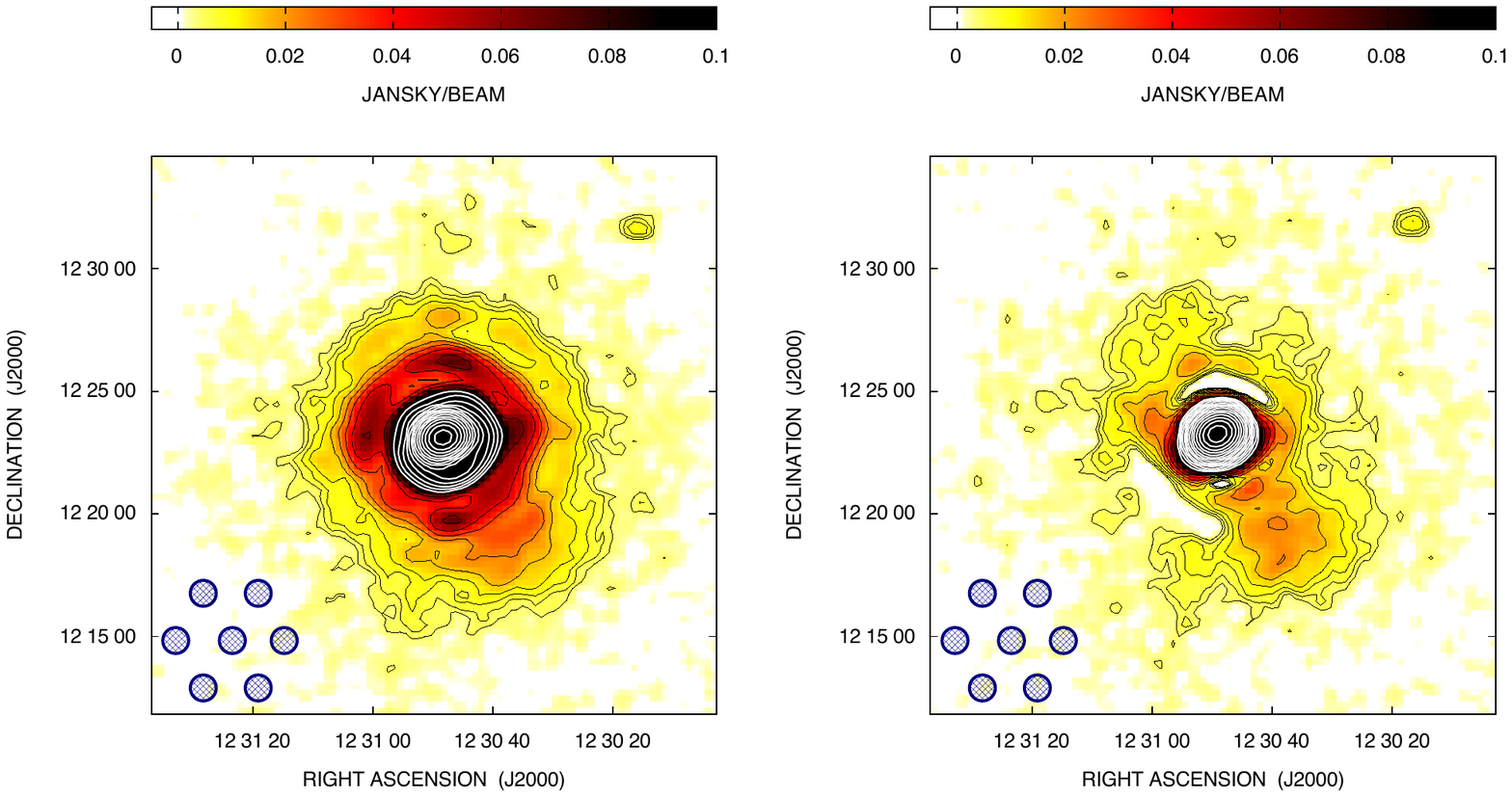}}
\vspace{-14.5cm}
\caption{{\it Left panel:} 'Dirty' image of Virgo A resulting from the combination of 10 OTF scans achieved with the multi-feed K-band array at a frequency of 19~GHz and a bandwidth of 2~GHz. Close to the source the dynamic range is limited to $DR\sim 30$, due to the strong secondary lobes.  {\it Right panel:} CLEANed image restored with a circular Gaussian with FWHM of $57^{\prime\prime}$. The measured surface brightness RMS sensitivity is now 1.3 mJy/beam and the dynamic range increases to $DR=11300$.  In both panels, contours start from $3\sigma$ and increase by a factor of $\sqrt{2}$. Due to the extreme DR the color map has been saturated to highlight the faint emission of the extended lobes of the radio source. In the bottom left corner of both panels the footprint on sky of the K-band multi-feed receiver is shown.  The size of each feed corresponds to the restoring beam.}
\label{fig:HDR-3C147}
\end{figure*}

\subsection{Dynamic Range and Deconvolution Techniques}\label{sec:HDR-3C147}
Image fidelity and dynamic range are usually primarily limited by secondary lobes of the antenna radiation pattern rather than by thermal noise 
fluctuations. In this section, we aim to demonstrate the SRT's imaging capabilities, in combination with deconvolution techniques, in the presence of a very bright point source that can severely limit the dynamic range. For this purpose, we chose to image Virgo~A, one of the most challenging sources in this respect, and we exploited the  SCUBE software package, which was optimized for SRT data reduction and analysis, and which includes deconvolution capabilities (see Sect.~\ref{sec:tools3} for more details).

Virgo~A is  the fourth brightest radio source in the northern sky and is associated with M87, the galaxy at the center of the nearest rich galaxy cluster (Virgo). 
The AGN hosted by this elliptical galaxy is powered by one of the most massive active black holes discovered so far (M$_{\rm BH} \simeq 6.4 \pm 
0.5 \times 10^9 $M$_{\sun}$, \citealt{gebhardt09}). For these reasons, it has been the subject of a large number of studies at all wavelengths, from 
the radio to the X-ray. In the radio band, it has been studied at low frequencies with interferometers like the VLA (\citealt{owen00}) and 
LOFAR (\citealt{degasperin12}). These observations reveal similar morphologies: the radio source is composed of a bright inner double, which 
contains a collimated one-side relativistic jet pointing towards the north-west that is embedded in a low surface brightness halo of 
synchrotron plasma. The angular size of the inner double is about 1 arcminute, corresponding to about 5 kpc in linear size. The faint halo 
extends for 15 arcmin (or 80 kpc). Despite its comparatively lower surface brightness, the extended halo is responsible for much of the flux density. 
Due to the large difference in surface brightness between the halo and the inner double, high-dynamic range imaging of Virgo A has always 
represented a big challenge.  At high radio frequencies ($\nu > 1$ GHz), interferometric 
observations are nearly impossible due to the large angular size of the source. On the other hand, large single dish antennas with low-noise receivers and large 
collecting areas, like the SRT, provide enhanced surface brightness sensitivity for the detection of faint large-scale radio emission and the 
recovery of the flux density missed by interferometers. Multi-wavelength observations and spectral studies (see \citealt{degasperin12}) can indeed provide important information about the relationship between the compact inner double and the extended halo in Virgo~A, which is still matter of debate.

Virgo A was observed through multi-feed OTF scans at K-band on 15 November 2015. 
Data were acquired with the TP backend at a central frequency of 19~GHz and a bandwidth of 2~GHz. We mapped a FoV of
 about $0.5^{\circ}\times 0.5^{\circ}$ with 121 sub-scans spaced by 15 arcsec on the sky using the de-rotator fixed in angle in the equatorial frame. We interleave the OTF scans with a telescope sky dip every hour to determine the trend of the optical depth $\tau$ during the observation and to correct the data to compensate for the atmosphere opacity ($\tau\simeq 0.05$ during these observations). The scanning speed was set to 6 arcmin s$^{-1}$, and we sampled at 10 ms. We acquired 10 OTF scans, 5 along the RA direction plus 5 along the DEC direction. We obtained useful data from 12 outputs of the multi-feed (7 from the left-circular polarization channel and  5 from the right-circular polarization channel), thus we collected a total number of 120 images centered on Virgo A. Multi-feed data reduction was performed with the SCUBE software package. We observed 2 OTF mapping of the primary calibrator 3C~286 that we used to convert the raw counts to the flux density scale of \citet{perley13} for all the seven feeds of the K-band array. We also corrected for variation of telescope gain with elevation using the curves derived during the AV (see Sect.~ \ref{sec:gaincurve}). 

Figure~\ref{fig:HDR-3C147} (left panel) shows the image obtained for Virgo A.
As explained in Sect.~\ref{sec:beam}, the asymmetries of the secondary lobes of the K-band beam patterns are strong, and limit the dynamic range to $DR\sim 30$ close to the source. Standard deconvolution techniques, like those implemented for radio interferometry data, can help  to remove the beam sidelobes and effectively reach the thermal surface brightness RMS sensitivity (expected to be $\sim 1$ mJy/beam for a pixel size of 15 arcsec, by applying Eq.~\ref{eq:noise}). In particular, it is necessary to perform a precise characterization of the telescope beam pattern followed by its accurate deconvolution from the dirty image of the sky brightness. Due to elevation-dependent asymmetries (see Sect.~\ref{sec:beam}), we need to employ an elevation-dependent beam model for a proper deconvolution of the sky image from the antenna pattern.

We used the wavelet beam model obtained on 3C 84 (see discussion on Sect.~\ref{sec:beam}) to deconvolve the antenna multi-feed beam pattern from the `dirty' image of Virgo A in the Equatorial frame using the SCUBE software.  The CLEAN algorithm rotates and interpolates the beam pattern model by taking into account the specific elevation and parallactic angle for each individual data point. 
For Virgo A, we measure a peak surface brightness of about 14.7 Jy/beam on the cleaned imaged. CLEAN components are convolved with a circular Gaussian with FWHM $57^{\prime\prime}$ (shown in the bottom-right corner of the right panel) and restored with the residual image. The resulting image is shown in the right panel of 
Figure~\ref{fig:HDR-3C147}.  The sidelobes are not an issue anymore, the noise is 1.3 mJy/beam (very close to the expected thermal noise) and the dynamic range increases to $DR=11,300$. The improvement obtained with the deconvolution is indeed very significant and allow us to reveal the faint and diffuse outer lobes of the radio source that were hidden below the secondary lobes of the beam pattern. 

Figure~\ref{fig:HDR-3C147} (right panel) represents the highest frequency observation ever achieved for Virgo A  at arcminute resolution, superseeding the image of the halo of Virgo A 
obtained at 10.55 GHz with the Effelsberg radio telescope at 69 arcsec resolution (\citealt{rottmann96}). At the SRT resolution, the inner double cannot be resolved, but our high-dynamic range observation has enough sensitivity to detect the full extent of the halo. 

The K-band multi-feed HDR imaging of Virgo A is perhaps the most complex imaging test performed during the AV, pushing the SRT to the limits of its present capabilities as a single-dish imaging instrument.  
This system is also targeted as part of one of  the early science SRT projects. For a full analysis of the 
high frequency spectral properties of Virgo A, we therefore refer to a future paper (Murgia et al., in prep.) where we report the results of this project. 

\begin{table}[t]
\caption{Observational parameters for the C-band observations of the Galactic extended sources discussed in Sects.~\ref{sec:HDR-SFR} and ~\ref{sec:IF}. }             
\centering 
\label{tab:obspar}   
\begin{tabular}{l c c c c}     % 3 columns 
\hline
Target  &  FoV &  $v_{scan}$ & $t_{int}$ & $t_{tot}$ \\  
             &    (deg$^2$) &  ($^{\prime}$/s) &      (ms)   &   (h)     \\
\hline\hline 
Omega Nebula  & 1$\times$1 & 3  & 10     & 6 \\
W3  & 2$\times$2 & 6  & 10  & 8  \\
IC443  & 1.5$\times$1.5 & 4  & 40 &   6.5 \\
\hline                  
\end{tabular}
\end{table}

\subsection{Dynamic Range: Extended Sources}\label{sec:HDR-SFR}
In the previous section, we dealt with the case of an extremely bright point source, which required the implementation of deconvolution techniques to achieve good image quality. Here  we want to determine the SRT dynamic range capabilities to map very large-scale, low-surface-brightness emission, filling up most of the field of view, with bright point sources embedded in it. In this case, deconvolution techniques are not essential, as the sidelobes produced by the bright point sources affect only a very limited portion of the image. For this purpose, we imaged two well-known Galactic extended sources: the Omega Nebula and the W3 molecular cloud complex.

The Omega Nebula and the W3 molecular cloud complex are considered to be among the brightest and most massive star-forming regions of our galaxy and are known to be very interesting examples of maser and radio recombination line sources (se OH and Methanol maser measurements presented in Sect.~\ref{sec:xarcos}).  The continuum radio emission from these objects is mostly due to free-free radiation emitted by Hydrogen ionized from the ultraviolet photons produced by young and massive stars. Both the Omega Nebula and W3 star-forming regions have been imaged as part of the  commissioning of the Green Bank Telescope (GBT). This means that  we can directly compare the imaging capabilities of two state-of-the-art instruments at similar observing frequency and resolution. 

We observed the two targets  in the period October-November 2014. We performed OTF maps along RA, DEC, GLON, GLAT, AZ, and EL directions. The data were acquired using the TP backend at a central observing frequency of 7.24 GHz and with a bandwidth of 680 MHz. The main observational parameters of both sources are listed in Table~\ref{tab:obspar}. For each target we list the imaged FoV, the mapping speed $v_{scan}$, the integration time $t_{int}$ (equal to the sampling time) and the total time $t_{tot}$ spent on each target. The results are discussed below.

\begin{figure}[t]
\centering
\vspace{0.3cm}
\resizebox{0.9\hsize}{!}{\includegraphics[angle=0]{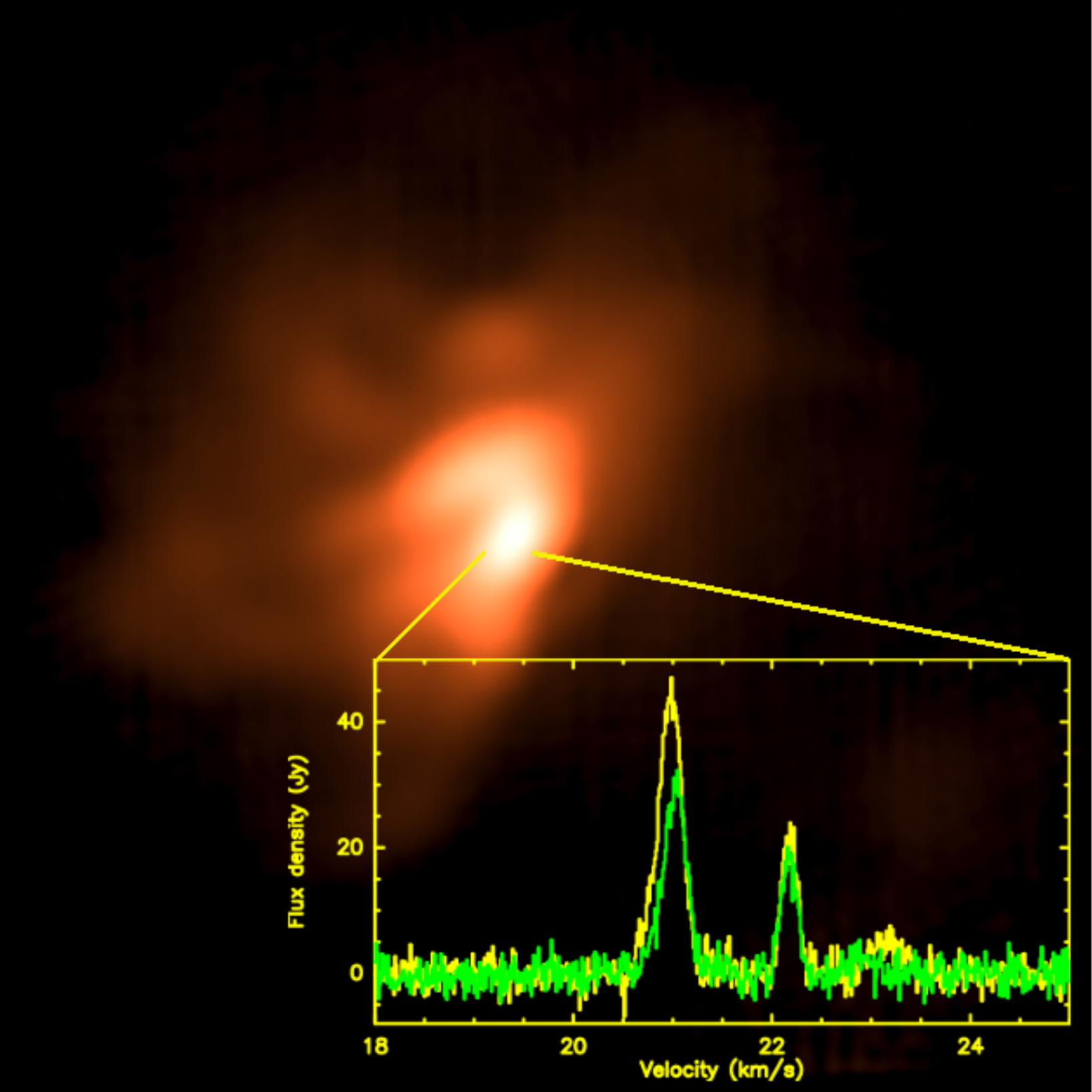}}
\vspace{0.7cm}
\caption{SRT continuum image of  the Omega Nebula at 7.24 GHz. The source emission shown extends over about a 1\deg$\times$1\deg~region.
The noise is $\simeq$7 mJy/beam and the SRT resolution  is 2.6 arcmin (SRT beam size at 7.24 GHz).
The colour map has been biased to highlight the low surface brightness emission. As a result, the central and brightest part of the nebula (M17-UC1), characterized by a surface brightness in the range $\simeq$ 3-100 Jy/beam, is saturated. 
{\it Inset spectrum: } 6035\,MHz OH maser emission (Jy) observed towards M17-UC1. Left and right circular polarizations are shown in  yellow and green, respectively. The spectrum was acquired with XARCOS, pointing at the position (J2000) R.A.$=18^{\rm h} 20^{\rm m} 24^{\rm s}.8$ and Decl.$=-16$\degr11\arcmin37\arcsec. The velocity scale refers to the LSR frame and uses the optical convention. The velocity resolution is $\Delta v=10$ m~$s^{-1}$ and the RMS noise level is $\sigma$=2 Jy/channel (for more details see Sect.~\ref{sec:xarcos}). }
\label{fig:Omega}
\end{figure}

\begin{figure}
\centering
\vspace{0.3cm}
\resizebox{0.87\hsize}{!}{\includegraphics[angle=0]{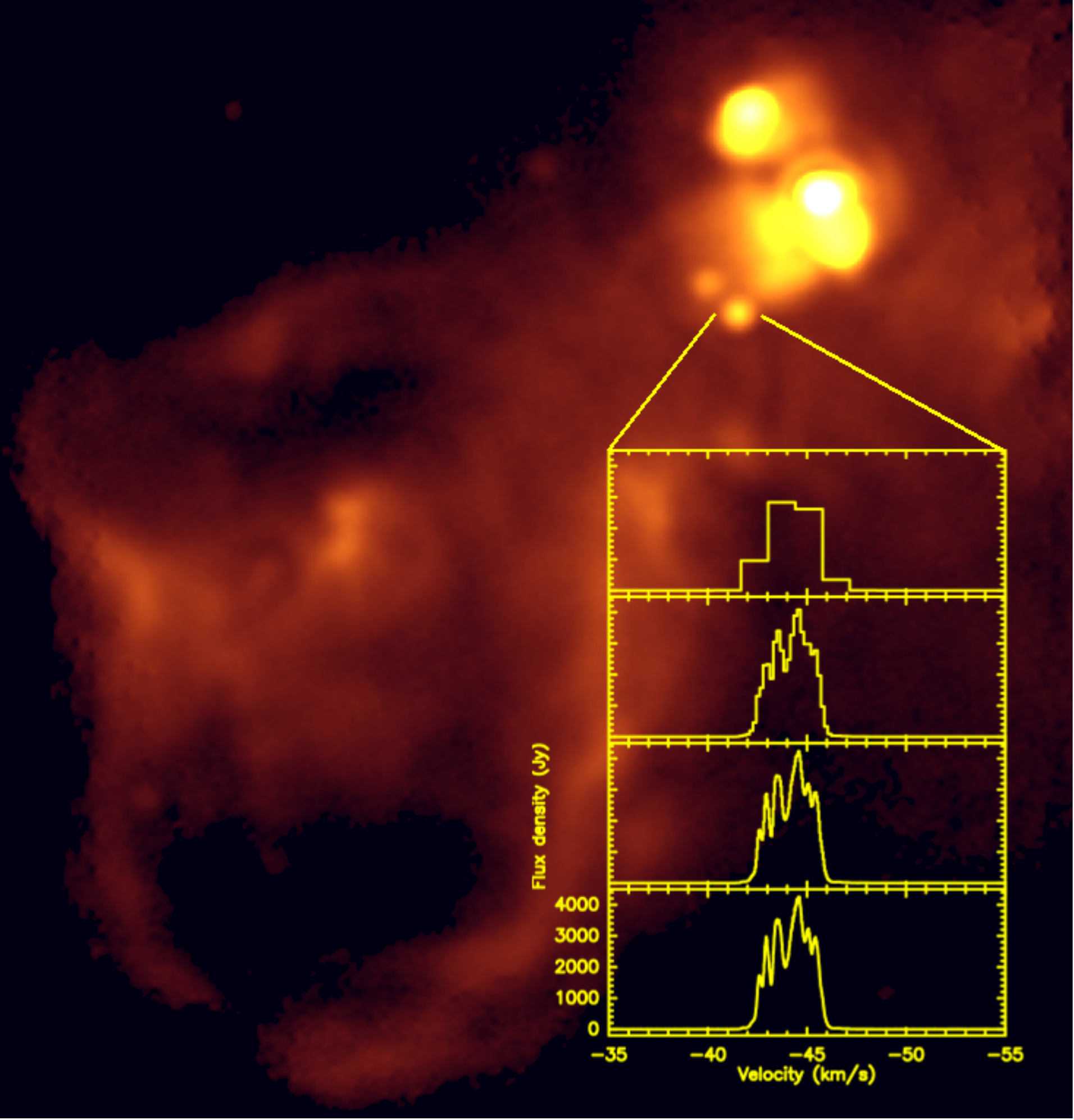}}
\vspace{0.7cm}
\caption{
 SRT continuum image of W3 at 7.24 GHz. The field of view is about 2\deg$\times$2\deg. The noise is $\simeq$4 mJy/beam and the  resolution is 2.6 arcmin (SRT beam size at 7.24 GHz). The colour map has been saturated to highlight the low surface brightness emission.  {\it Inset spectrum:} 6.7\,GHz methanol maser emission (Jy) in W3(OH) observed at the SRT  with XARCOS on January 28th, 2016. Four sub-bands (see panels) are observed simultaneously at increasing spectral resolution. From top to bottom: 1.4 km~s$^{-1}$, 170, 40, and 10 m~s$^{-1}$). Left and right circular polarizations have been averaged. The velocity scale refers to the LSR frame and uses the optical convention (for more details see Sect.~\ref{sec:xarcos}). 
}
\label{fig:W3}
\end{figure}

As for Virgo~A (see Sect.~\ref{sec:HDR-3C147}), the data of the Omega Nebula and the W3 molecular complex were reduced and analyzed using the SCUBE software. In this case, though, deconvolution techniques were not applied. We note that to evaluate the robustness of our surface brightness measurements, we also imaged two well known calibrators, 3C~295 and 3C~147, with an OTF mapping setup similar to that of the targets. These calibrators were analyzed through the same strategy  adopted for the Omega Nebula and W3, and the measured flux densities  (on the \citealt{baars77} scale) were checked for consistency against values from the literature.  

The image obtained for the Omega Nebula after flagging of RFI-corrupted scans, data calibration, baseline subtraction, imaging, and combination of the  six OTF maps obtained, is shown in Fig.~\ref{fig:Omega}. The nebula peaks at a surface brightness of 101.5 Jy/beam, and presents a northwest - southeast  orientation with a sharp edge on the west side. Two patches of extended emission are visible in the bottom right corner of the image. This is also the direction of the Galactic plane. 
The radio emission fades from the center of the nebula outwards, but it is clear that the faint extended emission covers the entire field of view of our observation. This limits the possibility of calculating the noise with a good precision. We evaluated a noise of $\simeq$7 mJy/beam. The resulting image dynamic range, measured as the peak to noise ratio, is DR$\sim 14500$. 

Figure~\ref{fig:W3} 
shows the image obtained for a field of view of about 2\deg$\times$2\deg, which includes the W3 main molecular cloud  (the brightest structure visible on the top right, which hosts the ultra-compact HII region W3)  and the low surface brightness  emission extending towards south-east, which is catalogued as IC~1805 (also known as the Heart Nebula). 
We evaluated a noise of $\simeq$4 mJy/beam. Considering that the peak of the emission associated with W3 Main reaches a surface brightness of 35.8 Jy/beam, the resulting dynamic range is DR$\simeq$9000. In addition to the thermal emission associated with the W3 and IC~1805 star forming region, we also observed a handful of point sources which are likely background extragalactic objects. 

\begin{figure*}[t]
\centering
\vspace{-2.5cm}
\resizebox{13cm}{!}{\includegraphics[angle=0]{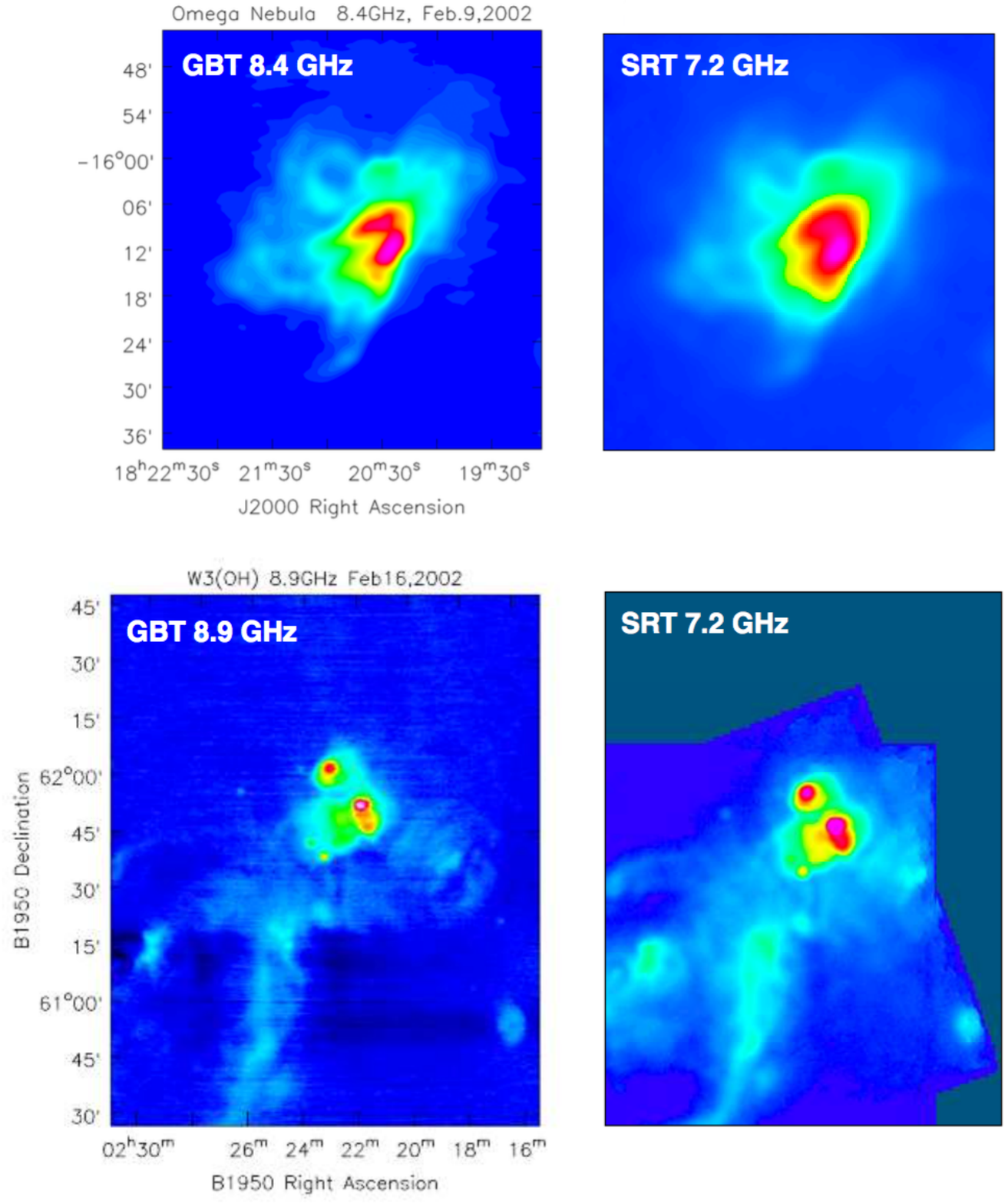}}
\hfill
\parbox[t]{50mm}{
\vspace{-6.2cm}
\caption[]{Comparison of GBT (left panels) and SRT (right panels) images of the Omega Nebula (top) 
and W3(OH) (bottom). The GBT commissioning observations of the Omega Nebula and of W3 were obtained 
at 8.4 GHz (\citealt{prestage03}) and  8.9 GHz (GBT site) respectively. }
\label{fig:gbt} }
\vspace{-3cm}
\end{figure*}

For a qualitative assessment of the imaging capabilities of the SRT in comparison with the GBT, we show  in Fig.~\ref{fig:gbt} the commissioning GBT images of the Omega Nebula 
(top-left panel) and W3 (bottom-left panel) obtained at 8.4 GHz and 8.9 GHz, respectively, along with the SRT images at 7.2\,GHz (top-right and bottom-right panels). For a better comparison, the SRT images have been rendered using a similar colour map, saturation, and field of view to the GBT images. The GBT images have an angular resolution that is better by a factor of two. As a consequence, the bright compact features appear better detailed in the GBT images. However, the SRT images recover all of the structures seen by the GBT and, in addition, the smoother resolution provides a better signal-to-noise ratio ($S/N$) for the faint and extended radio emission associated with the nebula. Indeed, we found that SRT could reach dynamic range levels  comparable to those reached with the GBT (i.e. $DR\sim 10000$), at least in the case of Omega Nebula. It is important to stress that the high dynamic range levels quoted above for extended sources have been obtained without deconvolution of the SRT beam pattern, In fact, the secondary lobes of the beam affect only the regions very close to the peak of the emission in these images. In these cases, the very accurate baseline subtraction allowed by the SCUBE software have proved to do an excellent job.

\begin{figure*}
\centering
\vspace{-2.0cm}
\resizebox{0.9\hsize}{!}{\includegraphics[angle=90]{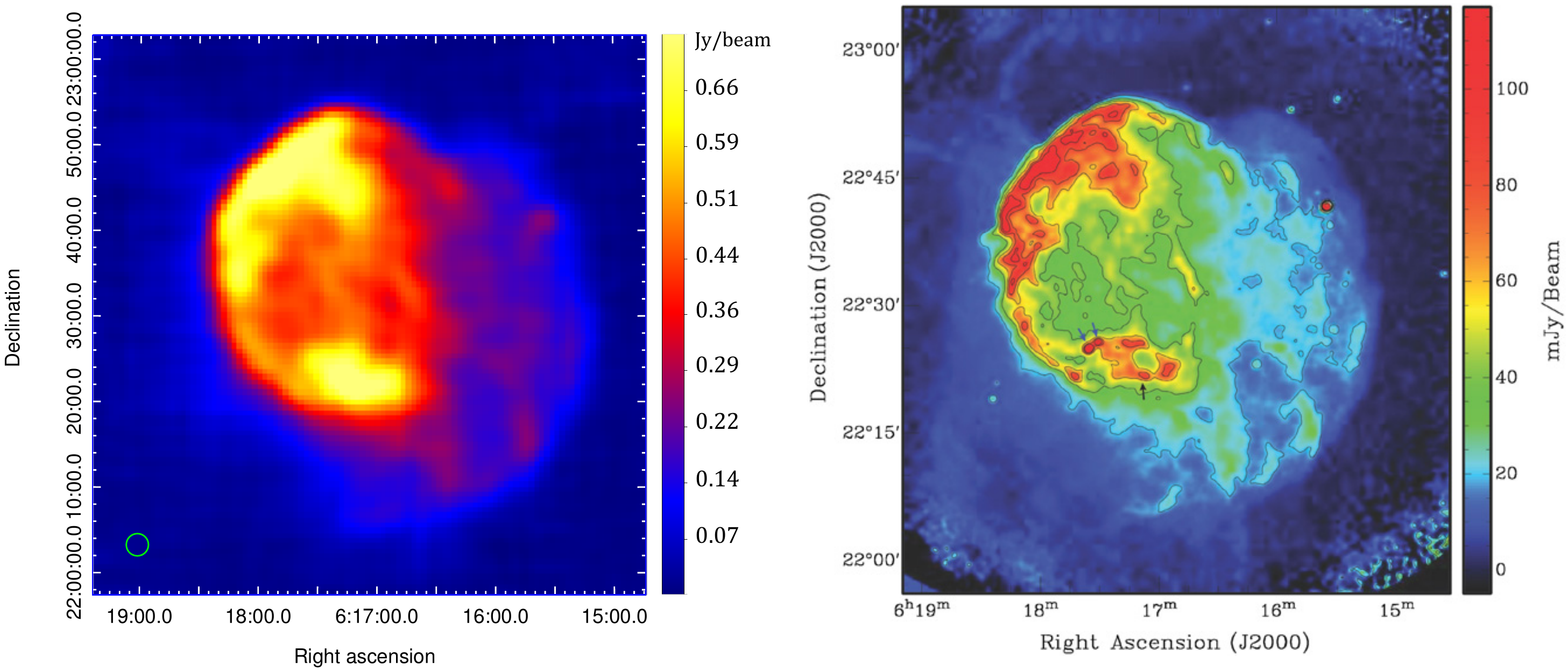}}
\vspace{-2.0cm}
\caption[]{Images of SNR IC443. {\it Left:} Single-dish 7.24 GHz calibrated image obtained with the SRT (2.7$^{\prime}$ resolution; image RMS $\sim 10$ mJy/beam for an effective exposure time of 6.5 hours). {\it Right:} 1.4 GHz continuum 7-pointing mosaic obtained with the  VLA (40$^{\prime\prime}$ resolution) in combination with Arecibo (3.9$^{\prime}$ resolution) data to provide sensitivity to extended low surface brightness emission (total observing time 6.3 hours). The blue arrows indicate the position of two extra-galactic background sources, and the black arrow indicates the neutron star (from \citealt{lee08}).}
\label{fig:3c157} 
\end{figure*}

\subsection{Image Fidelity: Galactic Plane Sources}\label{sec:IF}

The image fidelity capabilities  of the SRT were tested through observations of well-known SNRs. Among them,  IC443 represents an ideal target in this respect, thanks to its interesting complex morphology, to the availability of extensive multi-wavelength data, and to its good visibility at the SRT site. 

IC443 (also named 3C~157), located at about 1.5 kpc in the direction of the Galactic anticenter (\citealt{heiles84}; \citealt{braun86}), is one of the best-studied Galactic SNRs. The large structure of the source extends over 0.75\deg and shows evidence of interactions with both atomic and molecular clouds (Snell et al. 2005). The discovery of a neutron star in X-rays with Chandra (\citealt{olbert01}) suggests a core-collapse origin for the SNR. Its integrated flux density  at 1 GHz is about 160 Jy (for a detailed discussion of this source, see \citealt{green14} and references therein).  

We observed IC443 in the period May-December 2014. We performed OTF maps along RA and Dec directions. The data were acquired using the TP backend at a central observing frequency of 7.24 GHz and with a bandwidth of 680 MHz. The main observational parameters of the source are listed in Table~\ref{tab:obspar}. 
The adopted observing parameters typically imply the acquisition of $>$10-20 samples/beam for each subscan passage (largely oversampling the beam w.r.t. Nyquist sampling). This  allows us to get an accurate evaluation of flux density errors, as well as to efficiently reject outlier measurements ascribed to RFI.
The offset between two consecutive subscans was set to 0.01\deg, which implies on average 4.5 passages per beam, and about 17 samples per beam per scan (assuming a beam size of 2.7$^{\prime\prime}$ at our observing frequency). 
The total duration of an observation (single map RA+Dec) was about 3 hours including dead/slew time.

The length of the subscans (or the image size) was chosen according to the size of the source. In order to properly reconstruct the morphology of the observed source and its associated flux density, the subscan-dependent baseline (background) must be correctly subtracted. Ideally, each subscan should be free from significant source contribution (and RFI contamination) for 40-60\% of its length/duration, in order to properly identify and subtract the baseline component. This requirement is not trivially achieved for targets located in crowded regions of the Galactic Plane. 
The flux density calibration was based on cross-scan observations of six bright point sources: 3C147; 3C48; 3C123; NGC7027; 3C295; and 3C286. 

For data reduction and analysis, we used the SDI software, developed as part of the AV activities with the purpose to make it available to SRT users. 
 SDI provides a pipeline  for data inspection,  RFI rejection, baseline removal and standard image calibration, optimized for the SRT (see Sect.~\ref{sec:tools3} for more details).  While not allowing deconvolution (allowed by SCUBE),  SDI provides  innovative and accurate baseline subtraction techniques required for the imaging of crowded regions close to the Galactic Plane (see Pellizzoni et al., in prep. for more details).
An additional aim of  this test was therefore to assess the SDI performance and its robustness  in its automated operation mode. A calibrated SDI image obtained for SNR IC443 is shown in Figure~\ref{fig:3c157} where it is compared with VLA+Arecibo observations. 
The SRT image offers a detailed view  of the supernova remnant morphology, which is remarkably similar to that obtained through VLA interferometric observations carried out at lower frequency (1.4 GHz), and at a factor $2-4$ higher resolution. 
We estimated an image rms of $\sim 10$ mJy/beam and a dynamic range of $\sim$125. RFI typically affected $<$10\% of samples in the
adopted band and were automatically flagged by SDI.

The integrated flux density measured for IC443 is $64\pm 3$ Jy, consistent at a $1\sigma$ level with the expected value of $73\pm 10$ Jy, derived from the value of $84.6\pm 9.4$ Jy measured at 4.8 GHz with Urumqi (the sino-german 25m telescope, used to carry out a polarization survey of the Galactic plane) and assuming an overall spectral index of $-0.38\pm 0.1$ (\citealt{gao11}).  The main uncertainties in this flux comparison are the different angular resolutions of the two measurements (2.7$^{\prime}$ SRT and 9.5$^{\prime}$ Urumqi) and  the uncertainty on the source size, over which \citet{gao11}  integrated  the flux. 

For further details on C-band observations of SNRs with SRT and related early science outcomes, we refer to \citet{egron16b,egron16a}.

\section{Spectroscopy with XARCOS}\label{sec:xarcos}
At the end of 2015, the XARCOS multi-feed digital correlator  was fully integrated into the SRT control system {\it Nuraghe} (\citealt{melis15}). A number of  observations were then performed at the SRT to demonstrate its capabilities.  In particular, we performed observations of maser lines in a handful of well-known Galactic star forming regions. In the following, we report on spectral line (maser) observations taken at C-band on two targets: the Omega Nebula and W3(OH). 

Observations of both targets were performed on January 28th, 2016, using the standard position-switching mode. We used the mono-feed C-band receiver, which provides dual circular polarization. On-source and off-source scan integration times were equally set to 60 seconds. The off-position was shifted in Declination by two degrees from the target coordinates, to account for the large extension of the nebula. The XARCOS spectro-polarimeter was configured in the standard 'XC00' mode, which provides four bands of 62.5, 7.8, 2.0, and 0.5\,MHz, with 2048 channels for each band. This setup resulted in channel spacings of 1.5, 0.2, 0.05, and 0.01\,km~s$^{-1}$, respectively, for the four bands.

The Doppler correction for tracking the maser lines was performed by the FTrack software (Orlati et al., in prep.), which was fully implemented in the antenna control system by the time of our observations. The data reduction was performed using the GILDAS software, after the SRT FITS data were converted into GILDAS format using the `SRT\_CLASS\_WRITER' package (Trois et al. in prep.). 

\subsection{Omega Nebula} We observed the OH $^2 \Pi_{3/2}$, $J = 5/2$, $ F = 3 \rightarrow 3$ hyperfine transition, (rest frequency 6035.085\,MHz\footnote{We adopted the same rest frequency as used by \citet{knowles76}, to better compare our spectra to those shown in their Fig.~9 which have a similar spectral resolution (0.06\,km\,$s^{-1}$).}) towards the ultra-compact HII region M17-UC1 in the Omega Nebula. Each section was centered on the frequency corresponding to the $V_{\rm LSR}$ velocity of the strongest maser feature (21.0\, km\,s$^{-1}$) reported in \citet{knowles76}. We used the calibration mark and the gain curve presented in Sect.~\ref{sec:gaincurve} to establish the flux density scale of each polarization separately. We estimate that the accuracy of our absolute flux density scale is of the order of  $\sim$10\%. 

Excited-state OH maser emission at 6035\,MHz was first discovered towards M17 by \citet{rickard75} and confirmed by \citet{knowles76}, who detected strong emission features at $V_{\rm LSR}$=21.0\,km\,s$^{-1}$ and $V_{\rm LSR}$=22.1\,km\,s$^{-1}$. A third, more redshifted component was detected in later observations in the left polarization spectrum \citep{caswell95,fish06}. The line at $V_{\rm LSR}$=21.0\,km\,s$^{-1}$ was found to be partially elliptically polarized \citep{knowles76}. In addition, a separation of $\sim$0.04\,km\,s$^{-1}$ from the opposite circular polarization was measured for this line, which was attributed to Zeeman splitting in a +0.8\,mG  magnetic field \citep[where the positive sign indicates that the component of the magnetic field along the line of sight is pointing away from us;][]{fish06}. We detected all three components in the left circular polarization (LCP) spectrum of M17-UC1 and two in the right (RCP). The LCP and RCP spectra are shown in the  inset of Fig.~\ref{fig:Omega}. The results of Gaussian fitting to the line profiles are presented in Table~\ref{table:line_fit_omega}, where for each  component  we list the line velocity ($V_{\rm LSR}$), the line width ($FWHM$), the peak flux density ($S_{\rm peak}$), and the flux density integrated over the line ($\int S dV$). 

\begin{table}[t]
\caption{Line parameters of the excited-state OH maser towards the ultra-compact HII region M17-UC1 in the Omega Nebula. }             
\label{table:line_fit_omega}      
\centering        
\footnotesize
\begin{tabular}{c l l r r}     % 5 columns 
\hline\hline       
Pol. & \multicolumn{1}{c}{$V_{\rm LSR}$}     & \multicolumn{1}{c}{$FWHM$}          & \multicolumn{1}{c}{$S_{\rm peak}$} & \multicolumn{1}{c}{$\int S dV$}  \\
     & \multicolumn{1}{c}{(km\,s$^{-1}$)}   & \multicolumn{1}{c}{(km\,s$^{-1}$)}   &  (Jy)       & \multicolumn{1}{c}{(Jy km\,s$^{-1}$)} \\
\hline
LCP          & 20.974$\pm$0.002 & 0.289$\pm$0.004 & 44           & 13.5$\pm$0.2     \\
               & 22.177$\pm$0.003 & 0.186$\pm$0.006 & 22           & 4.4$\pm$0.1      \\
              & 23.14$\pm$0.02   & 0.34$\pm$0.03   & 5            & 1.9$\pm$0.2      \\
\hline
RCP          & 21.022$\pm$0.003 & 0.280$\pm$0.006 & 29           & 8.7$\pm$0.2       \\
             & 22.170$\pm$0.003 & 0.188$\pm$0.007  & 19           & 3.8$\pm$0.1      \\
\hline           
\end{tabular}
\end{table}

The line velocities are consistent with those measured by \citet{knowles76} (assuming a 0.1\,km\,s$^{-1}$ uncertainty in their measurements). 
We detected a Zeeman splitting of the strongest feature of 0.048$\pm$0.005\,km\,s$^{-1}$ that directly yields a magnetic field of +0.86\,mG, using the formula reported in \citet{baudry97}. Our result confirms the estimate made by \citet{fish06}, with a higher accuracy/confidence due to the higher spectral resolution of the XARCOS spectra.  

We caveat that Zeeman splitting determinations can be affected by systematics, like beam squint, that consists of the two circular polarizations (RCP and LCP) pointing to two slightly different
directions  (\citealt{heiles96}), and/or small amplitude non-Gaussian effects on the spectral  baselines (\citealt{vlemmings08}). Methods to calibrate these effects are 
discussed in, e.g., \cite{heiles01} and \cite{vlemmings08}. 
In this work, as in \cite{fish06}, we did not attempt to account for such systematics. 
\cite{vlemmings08} estimated a
contribution of beam squint to the measured magnetic field smaller than 40 $\mu$G 
in the case of W3(OH)  6.7-GHz methanol masers observations conducted with
the Effelsberg radio telescope. W3(OH) is a sort of worst-case
scenario because it is one of the most extended maser emission
sources, and beam squint effects are smaller when the maser region consists of individual compact
features that dominate the spectrum,  as for the case studied in this work.

Characterization of the SRT circular polarization performance for 
spectral - maser - line observations is currently on going, including the estimate 
of the beam squint discussed above. The characterization of the linear instrumental polarization, on the other hand,  has already been
successfully conducted and is presented in \cite{murgia16}. 

\begin{table}[t]

\caption{Line parameters of the CH$_3$OH maser in W3OH. }             
\label{table:line_fit_w3oh}      
\centering        
\footnotesize
\begin{tabular}{l l r r}     % 4 columns 
\hline\hline       
\multicolumn{1}{c}{$V_{\rm LSR}$}     & \multicolumn{1}{c}{$FWHM$}          & \multicolumn{1}{c}{$S_{\rm peak}$} & \multicolumn{1}{c}{$\int S dV$}  \\
\multicolumn{1}{c}{(km\,s$^{-1}$)}   & \multicolumn{1}{c}{(km\,s$^{-1}$)}   &  (Jy)       & \multicolumn{1}{c}{(Jy km\,s$^{-1}$)} \\
\hline
-45.44757$\pm$0.00003 & 0.54402$\pm$0.00008 & 2572 & 1399.1$\pm$0.5 \\
-45.05983$\pm$0.00001 & 0.1933$\pm$0.0002 & 1080   & 208.8$\pm$0.3 \\
-44.50744$\pm$0.00003 & 0.9251$\pm$0.0001 & 4123   & 3814$\pm$1 \\
-43.50565$\pm$0.00001 & 0.54010$\pm$0.00005 & 3436 & 1855.9$\pm$0.5 \\
-42.94927$\pm$0.00009 & 0.2627$\pm$0.0002 & 2696   & 708.2$\pm$0.6 \\
-42.5992$\pm$0.0001 & 0.3221$\pm$0.0006  & 1633    & 526$\pm$1 \\
\hline           
\end{tabular}
\end{table}

\subsection{W3OH}  We observed the (5$_1$--6$_0$) 6.7\,GHz CH$_{3}$OH maser transition (rest frequency 6668.52\,MHz) towards the famous ultra-compact HII region W3(OH), known to also host strong maser emission from several other molecular species (e.g., water and hydroxyl). Each section was centered on the frequency corresponding to the $V_{\rm LSR}$ velocity of the strongest maser feature (-45.1\, km) reported in \cite{menten91}.
As primary flux calibrator, we observed 3C286. The flux calibration was performed by estimating the average value of counts of the continuum in the on-off/off spectrum of 3C286 for the different sub-bands. 
This value ($\sim 0.14$ for all subbands and both polarizations), was used in conjunction with the expected flux density value of 3C286 at the observed frequency (6.11 Jy, computed following \citealt{baars77}), to convert counts into Jansky.

The main methanol maser line in W3OH is confidently detected in all four sub-bands. Figure~\ref{fig:W3}  (inset spectrum) shows the XARCOS  ability to provide simultaneous spectra at four different resolutions, from the  lowest resolution (30.5 kHz or 1.4 km~s$^{-1}$ at 6.7 GHz) spectrum covering a relatively broad frequency bandwidth  (62.5 MHz, top panel) to the highest resolution spectrum (240 Hz or 10 m~s$^{-1}$ at 6.7 GHz) over a narrow bandwidth (0.488 MHz, bottom panel). The former is particularly well suited to provide information on emission throughout the velocity field of the source, while the latter allows us to study line profiles in detail, revealing, for example, blue or red-shifted wings and/or complex line sub-components.  In particular, up to six features are revealed in the highest resolution spectra of the main methanol maser line in W3OH. A multi-Gaussian fit of the highest resolution W3(OH) spectrum was done utilizing the XGAUSSFIT routine of the FUSE IDL Tools\footnote{http://fuse.pha.jhu.edu/analysis/fuse\_idl\_tools.html}. The results are presented in Table~\ref{table:line_fit_w3oh} (columns are as in Table~\ref{table:line_fit_omega}; LCP and RCP spectra have been averaged in this case). The flux density of the peaks are consistent, within the uncertainties and taking into account possible variability, with those reported in the literature at comparable spectral resolutions (e.g., \citealt{menten91}). 

\section{Pulsar Observations }\label{sec:pulsar}

The SRT is currently equipped with two backends explicitly designed for pulsar applications: the ATNF Digital Filterbank DFB3 backend (Hampson \& Brown 2008\footnote{\tt http://www.jb.man.ac.uk/pulsar/observing/DFB.pdf}) and the ROACH board (see Sect.~\ref{sec:SRT} for more details). In Section~\ref{sec:pulsarC}, we report the results of preliminary test observations undertaken with the DFB3. DFB3 and ROACH  observations are compared  in Sect.~\ref{sec:pulsarL}, where we also discuss the pulsar timing performance of the SRT.

\subsection{Pulsar Observations at C-band}\label{sec:pulsarC}

\begin{table}[tbp]
  \centering
  \caption{Folding and search mode DFB3 configurations$(^*)$ tested for C-band observations. }
  \label{tab:DFBconf}
  \begin{tabular}{l | l}
    \hline\hline
    Folding mode & Search mode \\
    \hline
pdfb4\_1024\_1024\_1024 &  srch\_1024\_512 \\
pdfb4\_1024\_1024\_256 & srch\_512\_128 \\
pdfb4\_1024\_512\_512 & \\
pdfb4\_256\_1024\_512 & \\
pdfb4\_512\_1024\_1024 & \\
pdfb4\_512\_1024\_512 & \\
pdfb4\_512\_512\_512 & \\
    \hline
  \end{tabular}
\tablefoot{
\tablefoottext{*}{The three numbers reported for folding mode configurations refer to, in order: time bins in the pulse profile; bandwidth in MHz; and number of frequency channels. For search mode configurations only bandwidth and channels are indicated.}
}
\end{table}

\begin{figure*}
\begin{center}
\includegraphics[width=\textwidth,clip]{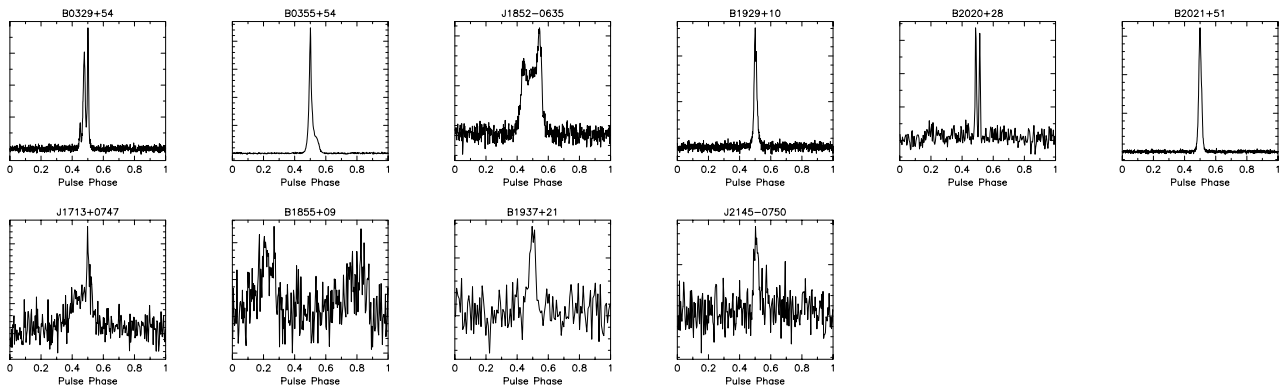}
\vspace{-8cm}
\caption[]{Pulse profiles for the pulsars observed at C-band with the
  DFB3. Long period, relatively bright pulsars are shown in the top
  row. 
  The four MSPs are shown in the bottom row. In both rows, the profiles are shown from left to right for increasing pulsar right ascension (i.e. in the same order as presented in Tab.~\ref{tab:PSRsC}).
  }
\label{fig:PSRsC}
\end{center}
\end{figure*}

The DFB3 backend performance was initially tested at C-band, as this was the lowest available frequency at the
time of our first test observations. C-band was preferred over K-band, as pulsars are expected to be brighter at lower frequencies, because of the negative slope of their power spectrum (e.g. \citealt{sieber73}).

The DFB3 produces files in psrfits format (\citealt{hotan04}), hence
SRT pulsar data can be directly handled with the most common, 
available pulsar software 
(e.g. psrchive\footnote{http://psrchive.sourceforge.net/},
dspsr\footnote{\tt http://dspsr.sourceforge.net/}, presto\footnote{\tt
  http://www.cv.nrao.edu/sransom/presto/}). The telescope code in the
file headers is {\tt SRT}. Most pulsar software packages have already been 
modified to include it; the tempo\footnote{\tt
  http://tempo.sourceforge.net/}/tempo2 (\citealt{hobbs06}) one letter
code for our telescope is {\tt z} and the alias is {\tt srt}.

C-band test observations started in June 2013 on a small number of
pulsars and continued until the installation of the L/P band coaxial
receiver.  The data acquisition was initially carried out using the
built-in TKDS graphical interface, which is now fully superseded by the new
data acquisition system named SEADAS (see Sect.~\ref{sec:tools2}).
The backend configurations  tested at C-band  are
listed in Table \ref{tab:DFBconf}. Both folding and search mode
observations, with sampling times down to 100 $\mu$s, were successfully
performed.

We observed six  bright long period pulsars and
 four MSPs for which a flux density measurement at a similar frequency was
reported in the literature along with a spectral index
(\citealt{kramer99}).   Table \ref{tab:PSRsC} lists the observed pulsars, along with the main observational details. For each observed pulsar,
    we report the number of observations performed ($N_{obs}$), the average length of each observation ($t_{obs}$),
   and  the average $S/N$ achieved. 
Figure \ref{fig:PSRsC} shows their integrated pulse profiles. 

\begin{table}[]
  \centering
  \caption{Pulsars observed at C-band. }
  \label{tab:PSRsC}
  \begin{tabular}{lcccc}
    \hline\hline
    PSR name & $N_{obs}$ & $t_{obs}$ & $S/N$ & S$_{6000}$   \\
     & &  (s) &  & (mJy)  \\
    \hline
%     & &   &  &  \\
 B0329+54 & 21 & 304 &124 & 2.5 \\
B0355+54 & 6 & 666 & 281 & 5.2 \\
J1852-0635 & 2 & 115 & 65 & 6.7 \\
B1929+10 & 3 & 164 & 122 & 4.6 \\
B2020+28 & 1 & 510 & 30 & 0.7 \\
B2021+51 & 6 & 303 & 244 & 4.2 \\
%     & &   &  &  \\
\hline
%     & &   &  &  \\
J1713+0747 & 1 & 1830 & 22 & 0.2 \\
B1855+09 & 1 & 930 & 13 & 0.4 \\
B1937+21 & 1 & 630 & 12 & 0.2 \\
J2145-0750 & 2 & 1365 & 10 & 0.1 \\
%     & &   &  &  \\
   \hline
  \end{tabular}
\tablefoot{
   In the top part of the table we list long period, bright
    pulsars while in the bottom part we list  millisecond
    pulsars. \\
    }
\end{table}

At the time of these observations, an accurate flux density and polarization calibration was not possible, as the noise diode (which allows the DFB3 to
directly switch its signal on and off) was not yet implemented through an optical link. 
An estimate of the pulsar flux densities at 6 GHz was then calculated from a modification of the radiometer equation for pulsars (see
e.g. \citealt{manchester01}). This estimate is reported in the last column of Tab.~\ref{tab:PSRsC}  ($S_{6000}$). 
Flux densities are broadly consistent (within a factor of $\sim$2-3)
with those found in the literature. This discrepancy is not
surprising given the inherent approximation of the uncalibrated method
used to estimate flux densities (that introduces errors $\ga 30\%$), and the fact that we used the (best case
scenario) tabulated value of the system temperature. Nevertheless, our C-band  observations allowed us to test and fine-tune the backend, as well as the pulsar acquisition and recording system.

\begin{figure*}
\begin{center}
\includegraphics[width=\textwidth,clip]{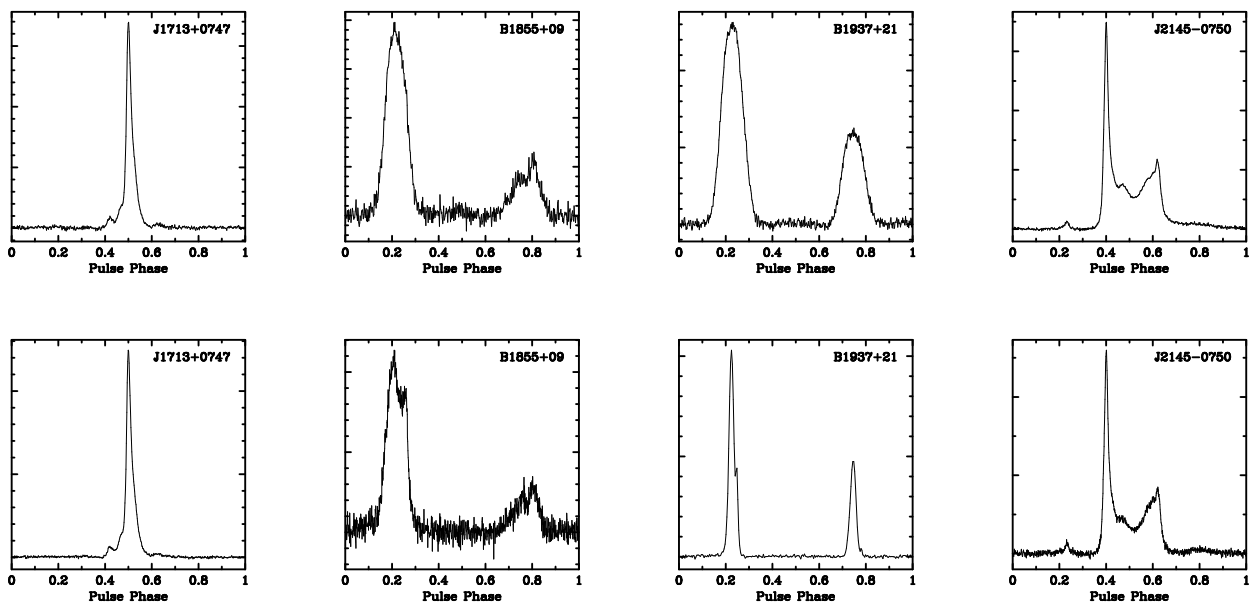}
\vspace{-5cm}
\caption[]{Pulse profiles at 1550 MHz for the four MSPs observed both
  at C and L-bands 
  The name of
  each pulsar is written on the top-right corner of each panel. The
  top row shows the pulses as observed with the DFB3, while in the
  bottom row we show those obtained with the ROACH board. The plots
  for the two backends are not obtained simultaneously, so the
 $S/N$ of the two profiles is not directly
  comparable. We also note that three of these pulsars (all
  except B1937+21) are heavily affected by interstellar scintillation.}
\label{fig:PSR_L}
\end{center}
\end{figure*}

\subsection{Pulsar observations at L-band}\label{sec:pulsarL}

\begin{table}[]
  \centering
  \caption[]{Main parameters of the four MSPs observed at both C- and L-band 
    with the DFB3 and ROACH backends. }
  \label{tab:PSR_L}
   \footnotesize
  \begin{tabular}{lrrr}
    \hline\hline
\multicolumn{1}{l}{PSR name} & \multicolumn{1}{c}{P0} & \multicolumn{1}{c}{DM}  & \multicolumn{1}{c}{S$_{1400}$}  \\
\multicolumn{1}{l}{} & \multicolumn{1}{c}{(ms)} & \multicolumn{1}{c}{(pc/cm$^3$) }  & \multicolumn{1}{c}{ (mJy) }  \\
\hline
 J1713+0747 & 4.57 & 15.99 & 10.20 \\
B1855+09 & 5.362 & 13.30 & $\cdots$  \\
  B1937+21 & 1.558 & 71.03 & 13.20  \\
  J2145-0750 & 16.052 & 9.00 & 8.90 \\
    \hline
  \end{tabular}
\tablefoot{All data are taken from
    the ATNF pulsar catalogue 
    (\citealt{manchester05}).\\
    }
\end{table}

When the L/P band coaxial receiver was installed and characterized, a
very extensive observing campaign started in which we exploited these
lower frequencies to observe $\sim$50 known pulsars  (taken from the ATNF pulsar catalogue\footnote{\tt http://www.atnf.csiro.au/people/pulsar/psrcat/}; \citealt{manchester05}) spanning a wide range of periods (from $\sim$1 ms to $\sim$1 s), dispersion measures 
(from $\sim$3 to $\sim$250 pc/cm$^3$) and 1400 MHz flux densities  (from $\sim$40 $\mu$Jy to $\sim$200 mJy) 
with all of the available DFB3 configurations for either folding or search observing  modes. A sub-sample was also observed with the ROACH board.

Since our backends have not yet been placed in a
screened chamber, and the control room itself is still in a
temporary location, the RFI situation at these lower frequencies is still
largely sub-optimal. In addition, an oscillating
noise diode was only recently installed for the L/P band receiver, and it is still under test. Despite these limitations, we were able to make 
extensive comparison tests of the  performance of the DFB3 and ROACH backends.

As test cases, here we present  the results obtained for the four MSPs that were also observed at
C-band. In Table \ref{tab:PSR_L} we present the main parameters of these MSPs: spin period (P0), dispersion measure (DM), and, when
    available, flux density at 1400 MHz (S$_{1400}$). In 
Figure \ref{fig:PSR_L}, we compare the pulse shapes
obtained with the DFB3 and with the ROACH board
at L-band (central frequency 1550 MHz). 

While the DFB3 has, at present, a
larger bandwidth (512 MHz vs 128 MHz for the ROACH), the ROACH board
allows us to coherently dedisperse the signal, fully correcting for
the dispersion delay. The DFB3, on the contrary, is limited by its
spectral resolution and cannot remove the intra-channel dispersion. In
most cases, this only has a minimal impact, or no impact at all, on
the observed profile. In the case of short period pulsars with a high
dispersion measure, though, the difference in the results obtained
with and without coherent dedispersion is significant. As we can
clearly see in the case of B1937$+21$, the intra-channel
dispersion smearing in the 1 MHz-wide channel of the DFB3
observation is, at the lower end of the observing band, $\sim 17$\%
of the spin period, which is much larger than the intrinsic pulse width
correctly recovered by the ROACH observation. 

\begin{figure}
\begin{center}
\vspace{-3.5cm}
\includegraphics[width=0.5\textwidth,clip]{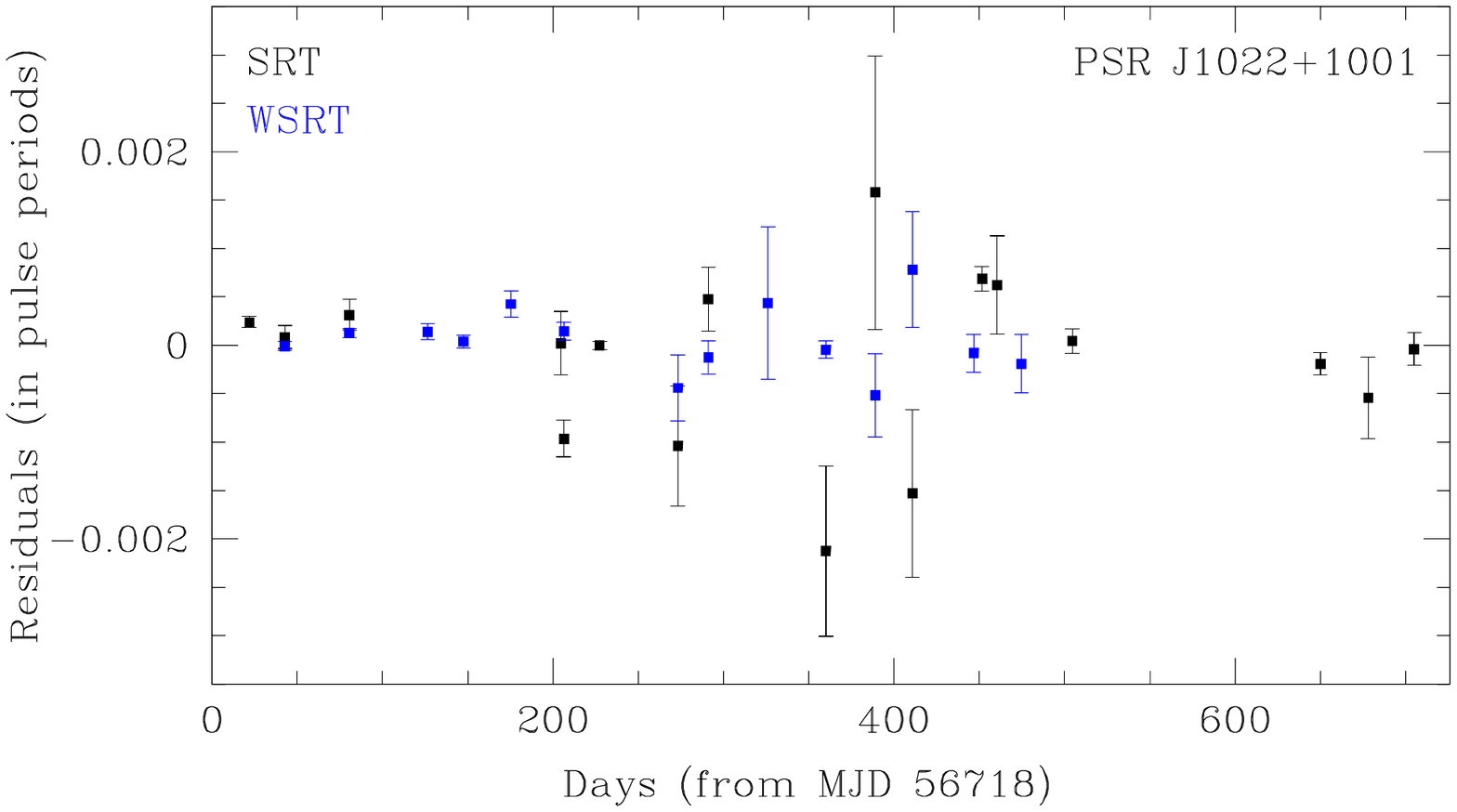}\\
\vspace{-3.7cm}
\caption[]{Timing residuals (observed minus predicted times of arrival
  vs time) for PSR J1022+1001, monthly observed for the LEAP project (Bassa
  et al. 2016) from March 2014 to January 2016. The black
  points refer to the SRT and the blue ones to WSRT. We note that
  the WSRT used 10 dishes in March and April 2014 (for an equivalent size
  of 75-m), nine from May to  August 2014 and eight from September 2014 to January 2015. The number of WSRT dishes available for LEAP observations decreased to seven  and then to five (66
  to 56-m equivalent size) from February to June 2015, when the last
  LEAP point at WSRT was taken (Janssen private communication). }
\label{fig:1022}
\end{center}
\end{figure}

To demonstrate the timing performance of the SRT, in Figure
\ref{fig:1022} we compare the times of arrival for the MSP J1022+1001 as
measured at the SRT (black) and WSRT (blue). These observations were taken as part of the LEAP
project (see Sect.~\ref{sec:LEAP} for more details), from March 2014 to January 2016. The measurements from the
two telescopes are consistent and the error bars are comparable (taking
into account the larger equivalent surface of the WSRT and the currently
worse RFI environment at the SRT), proving that the timing
capabilities of the SRT are up to standards. The increase in the error
bars at the beginning of 2015 (middle of the plot) is due to a
combination of effects: on one hand, the targeted pulsar appeared dimmer because
of scintillation; on the other hand, the WSRT was starting to use fewer
and fewer dishes for the LEAP project (see Figure \ref{fig:1022}
caption for details) and the SRT data suffered from packet loss, causing 8
MHz of bandwidth to become unusable for some of the epochs. After June
2015, WSRT stopped LEAP observations because of the upgrade of the
L-band receivers of the majority of the dishes to focal plane array
technology (Apertif; \citealt{verheijen08}), and only SRT measurements are available.

When the local RFI environment at the SRT is mitigated (by placing the equipment backends in a
Faraday cage and by screening the signals produced by the equipment in
the gregorian focus), we expect an
improvement of a factor of at least two in the $S/N$  of
our observations. 

A  more complete characterization of
the full potential of the SRT for pulsar observations will be presented in
a future paper, as soon as the on-site RFI situation is better managed, and the oscillating noise diodes at L- and P-bands become fully operational.

\section{LEAP Observations}\label{sec:LEAP}

Starting in July 2013, the SRT joined the LEAP project, a sub-project of the EPTA, performing observations of a set of millisecond pulsars simultaneously with the other four radio telescopes of the collaboration, using the L-band receiver and the ROACH backend. 
The pulsar timing data are reduced locally and the raw baseband data are shipped to Jodrell Bank, where the baseband data from all individual telescopes are correlated to find the phase offsets between the telescopes. The data are then corrected for the offsets and added coherently to form a tied-array beam, obtaining in this way high precision timing data. The high precision timing that is possible with LEAP will in turn be used, over timescales of a minimum of 5-10 years, to search for gravitational wave signatures. The details of the LEAP project and first results are shown in \cite{bassa16}. 

\begin{figure}
\begin{center}
\includegraphics[width=0.42\textwidth]{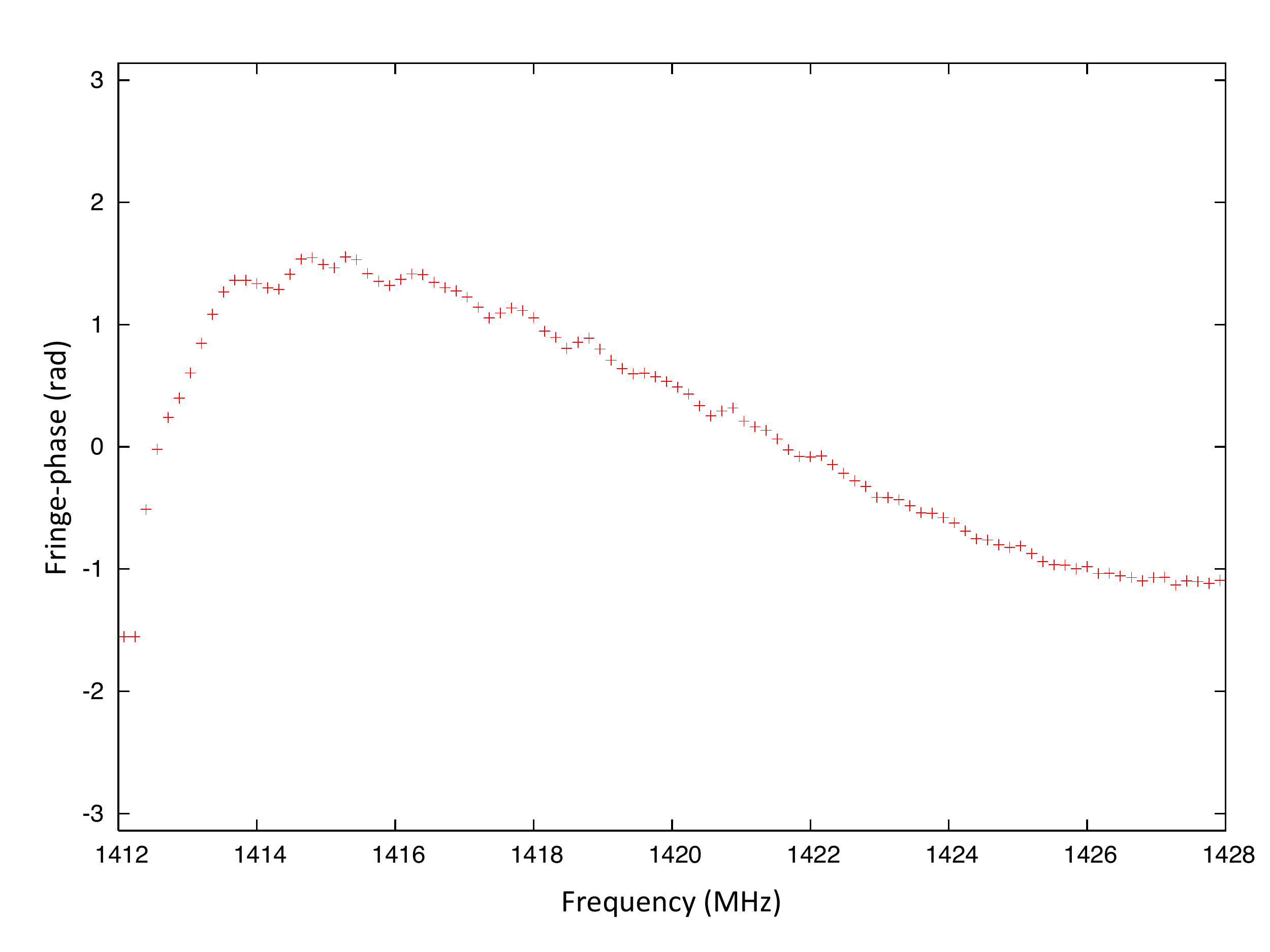}\\
\caption[]{First fringe obtained between SRT and WSRT using the LEAP correlation pipeline: 5-s long observation of the quasar 3C454 at 1420 MHz (May 2014). }
\label{fig:fringe}
\end{center}
\end{figure}

The SRT initially observed in a single 16-MHz sub-band of the LEAP bandwidth (1332-1460 MHz); starting in the spring of 2014, it has performed monthly observations in the full LEAP bandwidth and with the full set of millisecond pulsars (thanks to an 8-node CPU cluster and storage computer being installed on-site; see \citealt{prandoni14,perrodin14,perrodin16} for more details about the setup). In May 2014, we found the first fringe between the SRT and another telescope of the collaboration, the WSRT, using the LEAP correlation software at Jodrell Bank, thus showing the successful addition of the SRT to the LEAP tied-array. This is shown in Figure \ref{fig:fringe}.

\begin{figure}
\begin{center}
%\vspace{-0.5cm}
\includegraphics[width=0.32\textwidth,angle=-90]{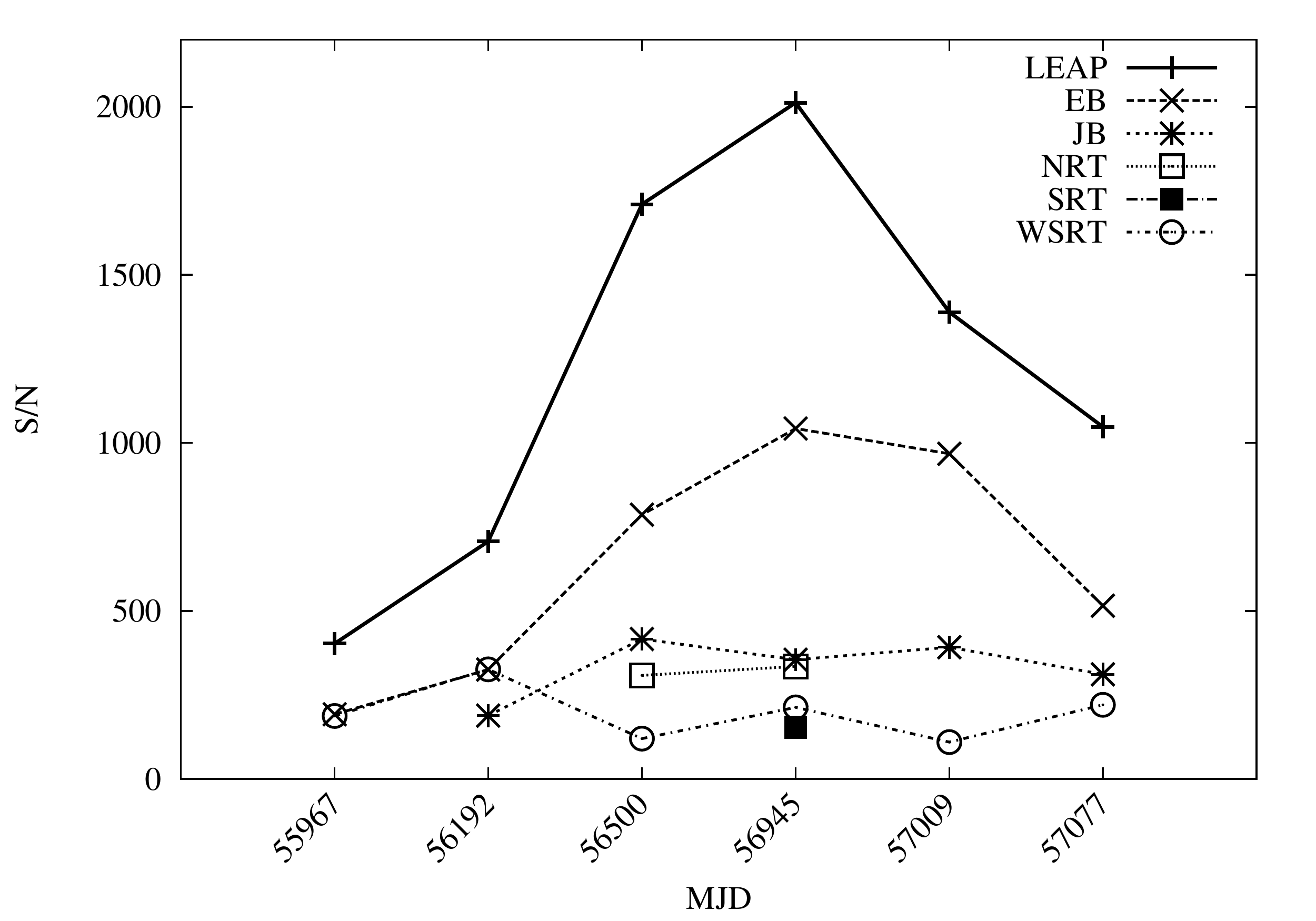}
\caption[]{$S/N$ from LEAP coherent additions as compared to the $S/N$ from the individual telescopes for PSR J1022+1001. The observations span the period from February 2012 to February 2015. We note that PSR J1022+1001 is affected by scintillation and its $S/N$  can vary wildly between observations. The SRT is included in MJD 56945.}
\label{fig:bassa}
\end{center}
\end{figure}

We have now collected over two years of monthly LEAP data at the SRT, improving in this way the $S/N$ of the LEAP data set. In coherent additions, the $S/N$ of pulsar timing signals should increase linearly with the number of telescopes, as opposed to the square root of the number of telescopes in the case of incoherent additions. Of course this is valid if the telescopes are identical in size and performance; in the case of LEAP, the $S/N$ of the coherent sum is expected to be equal to the sum of the $S/N$ of the individual telescopes. In Figure \ref{fig:bassa}, we show LEAP coherent sums for PSR J1022+1001 (from \citealt{bassa16}). The SRT is included in one epoch (MJD 56945), where it contributes roughly the same level of $S/N$ as the WSRT to the combined LEAP $S/N$. This confirms the comparison between SRT and the WSRT which was shown in Figure \ref{fig:1022}. We expect that the $S/N$ of LEAP observations at SRT will continue to
increase as most sources of RFI are identified and mitigated.

\section{Target of Opportunity Observations}\label{sec:ToO}

Since 2013, the SRT has been made available to the international community for 
ToO observations on a best effort basis, as part of the AV activities. Primary targets of SRT ToOs have been transient magnetars after X-ray
outbursts and Fast Radio Bursts (FRB).   

\begin{figure} \begin{center}
\includegraphics[width=0.5\textwidth,clip]{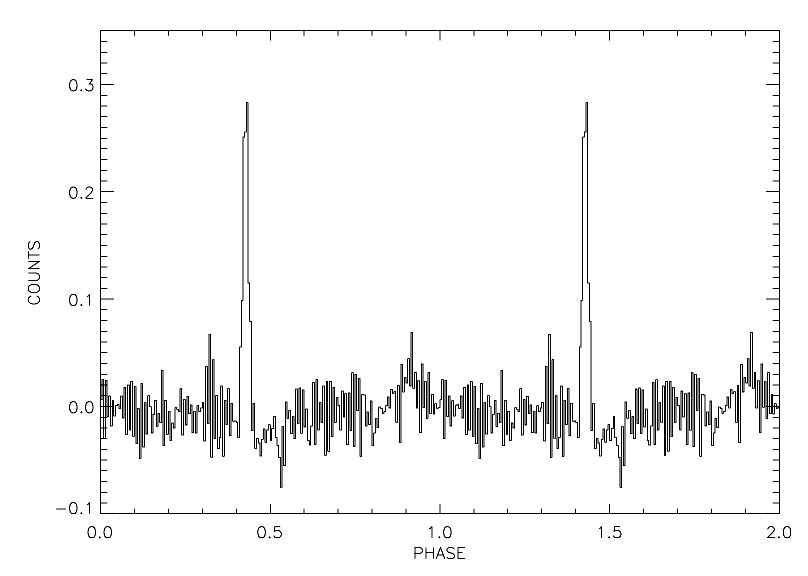} \caption{Pulse
profile at 7.3 GHz for SGR J1745-2900 folded with the period obtained
by Swift (\citealt{Gotthelf13}) and refined to maximise the
$S/N$.}
\label{fig:SGR1745}
\end{center}
\end{figure}

The ToO programme was kicked off on May 6th, 2013, before its
official opening, internally by the AV team, with observations of
the transient magnetar SGR J1745-2900. This peculiar source, showing a
3.67 s period pulsation, was discovered only a few days earlier by the
NuSTAR X-ray telescope in the vicinity of the Galactic centre (\citealt{mori13}). Despite the fact that the L/P band receiver and
the DFB3 and ROACH backends designed for pulsar observations were not yet installed
at the time, the SRT successfully detected radio pulsations from the
magnetar for the first time at upper C-band (\citealt{buttu13})
through observations performed with the TP backend, sampled
every 40ms over a bandwidth of 680 MHz. This was possible thanks to
the peculiar characteristics of radio magnetars (long periods and
roughly flat radio spectra) and to the prompt development (by members of the AV team) of an ad-hoc software tool able to fold the Total
Power data using the X-ray rotational ephemerides for the magnetar and
to optimize the spin period to obtain maximum $S/N$ for
the pulsed profile (Fig. \ref{fig:SGR1745}). This 7.3 GHz observation,
whose estimated flux density is 0.1 mJy (\citealt{buttu13}), nicely complements the detections at 8 and 5 GHz performed at
the Effelsberg radio telescope, where radio pulsations of the magnetar
were first discovered (\citealt{eatough13a,eatough13b}).

Other examples of  transient magnetars observed at the SRT are  the
magnetar candidate SGR J1819-1600 (\citealt{mereghetti12,page15}), and the Soft Gamma-ray Repeaters SGR
J1935+2154 (\citealt{israel14}) and SGR J0755-2933
(\citealt{barthelmy16}). 

A  {\it Letter of Intent} was signed on January 13, 2015, which included the SRT in the follow-up observations of FRB discoveries from the SUPERB (SUrvey for Pulsar and Extragalactic Radio Bursts) project (Keane et al., in prep.).
As part of this agreement, the SRT 
performed L-band and P-band follow-up observations of FRB150418 (\citealt{keane16}), FRB151206, FRB151230, and FRB160102
(Bhandari et al., in preparation), promptly after receiving a trigger
from the SUPERB collaboration. With the detection by \cite{spitler16} of repeating occurrences of pulses from FRB121102 (\citealt{spitler14}), further observations of FRB151206,
FRB151230, and FRB160102 were performed at L-band and P-band, while  
FRB121102 was observed at C-band for six hours. No single pulses
around the discovery dispersion measure of the FRBs were detected in
the data. 

The main observational details of  both SGRs and FRBs observed at the SRT as ToOs are listed in Table \ref{tab:too}. 
For each source, we list the date of the observation (in
  YYYYMMDD format), central observing frequency in GHz ($\nu_{obs}$), total bandwidth in MHz (BW),
  frequency resolution in MHz ($\Delta \nu$), time resolution in $\mu$sec ($t_{sampl}$), backend
  used, and observation length ($t_{obs}$). The last column lists the measured flux
  density or upper limit ($S$; in mJy for SGRs and in Jy for
  FRBs). For FRB flux density estimates, we explicitly indicate the dependence on the pulse duration $W$ (normalized over the actual duration of the observed pulse, in ms). This allows us to take into account pulse width variations that may occur in following burst episodes.

Several other ToOs have been requested in the past years on a variety of
types of astronomical objects, from comets to ultra-luminous X-ray
sources to GRBs (see e.g. the GRB~151027A, \citealt{nappo16}). A full description of such observations is beyond the scopes of this paper, and we refer the reader to the literature for more details.  

\begin{table*}[htbp]
  \centering

  \caption{Main observational parameters for SGR (top) and FRB (bottom)
  ToOs at the SRT.  }

  \label{tab:too}
  \begin{tabular}{ccccccccc}
\hline\hline
Source & Date & $\nu_{obs}$  & BW  &  $\Delta\nu$ & $t_{sampl}$  & Backend &
$t_{obs}$  & $S^{(\dag)}$ \\
 &  &  (GHz) &  (MHz) &  (MHz) &  ($\mu$s) &  &
 (s) &  (mJy) \\
\hline
SGR J1745-2900 & 20130506 & 7.3 & 680 & 680 & 40000 & TP & 7200 & 0.1 \\
SGR J1935+2154 & 20140731 & 6.4 & 412 & 1 & 6338 & DFB & 5598 & < 0.08 \\
               & 20140807 & 0.35 & 105 & 0.25 & 3169 & DFB & 1800 & \dots \\
SGR J1819-1600 & 20151126 & 1.55 & 500 & 2 & 250 & DFB & 8424 & < 0.07 \\
               & 20151126 & 6.5 & 924 & 2 & 500 & DFB & 7835 & < 0.04 \\
               & 20151206 & 1.55 & 500 & 4 & 250 & DFB & 5500 & < 0.08 \\
               & 20151206 & 6.6 & 924 & 2 & 650 & DFB & 5500 & < 0.05 \\
SGR J0755-2933 & 20160319 & 6.6 & 924 & 2 & 250 & DFB & 3766 & < 0.06 \\
\hline\hline
Source & Date & $\nu_{obs}$  & BW  &  $\Delta\nu$ & $t_{sampl}$  & Backend &
$t_{obs}$  & $S^{(\dag)}$  \\
 &  &  (GHz) &  (MHz) &  (MHz) &  ($\mu$s) &  &
 (s) &  (Jy) \\
\hline
FRB150418 & 20150421 & 1.55 & 500 & 2 & 125 & DFB & 3534 & $< 0.54\times(W/3.5)^{-0.5}$ \\
          & 20150421 & 0.35 & 105 & 0.25 & 128 & ROACH1 & 4363 & \dots  \\
FRB151206 & 20151207 & 1.5 & 500 & 1 & 500 & DFB & 12177 & $< 0.31\times(W/2.7)^{-0.5}$ \\
          & 20151207 & 1.5 & 500 & 1 & 500 & SARDARA & 10126 & \dots  \\
          & 20160506 & 1.5 & 500 & 1 & 250 & DFB & 10480 & $< 0.31\times(W/2.7)^{-0.5}$\\
          & 20160506 & 0.35 & 105 & 0.25 & 256 & ROACH1 & 10480 & \dots  \\
FRB121102 & 20160309 & 6.6 & 924 & 2 & 250 & DFB & 19792 & $< 0.07\times(W/2.8)^{-0.5}$\\
FRB151230 & 20160510 & 1.55 & 500 & 1 & 250 & DFB & 10350 & $< 0.24\times(W/4.6)^{-0.5}$ \\
FRB160102 & 20160507 & 1.55 & 500 & 1 & 250 & DFB & 7200 & $< 0.28\times(W/3.3)^{-0.5}$ \\
FRB160102 & 20160507 & 0.35 & 105 & 1 & 256 & ROACH1 & 7200 & \dots \\
\hline
  \end{tabular}
 \tablefoot{ 
 \tablefoottext{\dag}{For L-band observations the flux density was estimated by multiplying the nominal
  T$_{\rm sys}$ by two, to roughly take into account the  adverse RFI environment. P-band observations were too badly affected by RFI and a reliable flux  density estimate could not be obtained. For FRB flux density estimates we explicitly indicate the dependance  on the pulse duration $W$, normalized over the actual width of the observed pulse (expressed in ms). }
  }
\end{table*}

\section{Summary}\label{sec:summary} 

In this paper, we have provided a brief overview of the SRT and of the science applications envisaged  for it. We then  reported the main results of its astronomical validation, which was carried out in the period 2012--2015. 

The astronomical validation  activities started during the technical commissioning of the telescope, when several external software tools were developed, in order to assist the preparation, execution and monitoring of the observations, as well as the data inspection and reduction. This suite of astronomer-oriented software tools are meant to support future observers on-site. 

The scientific commissioning of the SRT, based on first-light receivers and backends,  then proceeded in steps, from basic  "on-sky" tests  aimed at verifying the general performance and/or the limits of the telescope and the acquisition systems, to more complex acquisitions aimed at assessing the actual SRT capabilities  for a range of scientific observations. 

The  activities were prioritized based on technical readiness and scientific impact. Highest priority was given to make the SRT available for joint observations as part of European networks: the European VLBI Network; and the Large European Array for Pulsars. 

After the first successful VLBI  data correlation of the Medicina--SRT baseline in January 2014 and the following  EVN test observations (see \citealt{prandoni14}), the SRT was offered as an additional EVN station in shared-risk mode for all of its three first-light observing bands (L-, C- and K-bands), starting from 2015. 

In parallel, all of the hardware and software necessary for the LEAP project was developed and installed at the SRT, where it was fully tested and debugged.  Since early 2014, the SRT participates in monthly LEAP runs, for which data acquisition is now fully automated. The first results of the LEAP project, including the SRT, are presented in \citet{bassa16}.

The validation of single-dish operations for the suite of SRT first light receivers and backends continued until the end of 2015, when the SRT capabilities for imaging, spectroscopy, and pulsar observations were demonstrated. In this paper, we have highlighted in particular the superb imaging performance of the SRT, both in terms of dynamic range and image fidelity.

Since 2013,  a ToO programme is offered to the
community on a best-effort basis.  The ToO programme was kicked off with the follow-up observation of
the transient magnetar SGR J1745-2900,  detected at the SRT at 7.3 GHz (\citealt{buttu13}). In January 2015,  the SRT was included in the agreement for follow-up observations of fast radio burst discoveries from the SUPERB  project (Keane et al., in prep.).
A formal agreement has also been established to include the SRT in a large multi-wavelength program, aimed at identifying and initially following up the electromagnetic counterparts of gravitational wave events.

The astronomical validation activity was formally concluded with the first call for shared-risk/early-science observations issued at the end of 2015. Early-science observations started on February 1st, 2016, and were conducted for a period of six months. These observations are aimed at further demonstrating the scientific potential of the SRT, and several papers are expected to appear in the literature in the coming months. 

\begin{acknowledgements}
The Astronomical Validation activities were made possible thanks to the invaluable support of the local technical staff and of the technical commissioning team. 
The authors thank Cees Bassa and Ramesh Karuppusamy for their important help in
setting up the ROACH backend for pulsar observations. D.~Perrodin thanks Gemma Janssen for providing WSRT timing residuals shown in
Figure \ref{fig:1022}. 
F.~Loi gratefully acknowledges the Sardinia Regional Government for 
financial support of her PhD scholarship (P.O.R. Sardegna F.S.E.
Operational Programme of the Autonomous Region of Sardinia, European
Social Fund $2007-2013$ - Axis IV Human Resources, Objective l.3, Line of
Activity l.3.1.). 
From 2012 to 2014, D.~Perrodin was supported by the ERC Advanced Grant
``LEAP", Grant Agreement Number 227947 (PI M. Kramer), which also provided a
10-gigabit-ethernet switch and a storage cluster for SRT LEAP observations.
M. Pilia was supported by the Sardinia Regional Government through the project "Development of a Software Tool for the Study of
Pulsars from Radio to Gamma-rays using Multi-mission Data" (CRP-25476). 
A. Ridolfi and C. Tiburzi gratefully acknowledge support from the
Max-Planck-Institut f\"ur Radioastronomie. C.Tiburzi also acknowledges support from
the Universitaet Bielefeld.
V.~Vacca was supported by the  DFG  Forschengruppe 1254 "Magnetisation of Interstellar and Intergalactic Media: The Prospects of 
Low-Frequency Radio Observations". 
The development of the SARDARA backend has been funded by the Autonomous 
Region of Sardinia (RAS) using resources from the Regional Law 7/2007 
"Promotion of the scientific research and technological innovation in 
Sardinia" in the context of the research project CRP-49231 (year 2011): 
"High resolution sampling of the Universe in the radio band: an 
unprecedented instrument to understand the fundamental laws of the 
nature". The Sardinia Radio Telescope  is funded by the Department of University and Research (MIUR),
the Italian Space Agency (ASI), and the
Autonomous Region of Sardinia (RAS). It is operated as national facility by the National Institute for Astrophysics (INAF).

\end{acknowledgements}

\bibliographystyle{aa} % style aa.bst
\bibliography{prandoni_SRT_ref}

%-------------------------------------------------------------------

\end{document}